\documentclass{aa}  

\usepackage{graphicx}
\usepackage{txfonts}
\newcommand{\teff}{\ifmmode T_{\rm eff} \else $T_{\mathrm{eff}}$\fi}
\newcommand{\logg}{\ifmmode \log g \else $\log g$\fi}
\newcommand{\lL}{\ifmmode \log \frac{L}{L_{\odot}} \else $\log \frac{L}{L_{\odot}}$\fi}
\newcommand{\mdot}{$\dot{M}$}
\newcommand{\myr}{M$_{\odot}$ yr$^{-1}$}
\newcommand{\vsini}{\ifmmode v \sin i  \else $v \sin i$ \fi}
\newcommand{\vinf}{\ifmmode v_{\infty} \else $v_{\infty}$ \fi}

\newcommand{\vmac}{v$_{\rm mac}$}

\newcommand{\kms}{\ifmmode $km~s$^{-1} \else km~s$^{-1}$\fi}
\newcommand{\msun}{\ifmmode M_{\odot} \else $M_{\odot}$\fi}
\newcommand{\zsun}{\ifmmode Z_{\odot} \else $Z_{\odot}$\fi}
\newcommand{\lsun}{\ifmmode L_{\odot} \else $L_{\odot}$\fi}
\newcommand{\Rsun}{\ifmmode R_{\odot} \else $R_{\odot}$\fi}
\newcommand{\qh}{\ifmmode Q_{\rm H} \else $Q_{\rm H}$\fi}
\newcommand{\qhei}{\ifmmode Q_{\ion{He}{i}} \else $Q_{\ion{He}{i}}$\fi}
\newcommand{\mum}{\ifmmode \mu $m$ \else $\mu$m\fi}

\begin{document}

   \title{Evolutionary status of the Of?p star HD\,148937 and of its surrounding nebula NGC\,6164/5\thanks{{\it Herschel} is an ESA space observatory with science instruments provided by European-led Principal Investigator consortia and with important participation from NASA.}\fnmsep\thanks{Based in part on observations collected at the European Southern Observatory, in Chile.} }


   \author{L. Mahy
          \inst{1}\fnmsep\thanks{F.R.S-FNRS Postdoctoral researcher}
          \and
          D. Hutsem{\'e}kers\inst{1}\fnmsep\thanks{F.R.S-FNRS Senior Research Associate}
          \and
          Y. Naz{\'e}\inst{1}\fnmsep\thanks{F.R.S-FNRS Research Associate}
          \and
          P. Royer\inst{2}
          \and
          V. Lebouteiller\inst{3}
          \and
          C. Waelkens\inst{2}
          }

   \institute{Institut d'Astrophysique et de G{\'e}ophysique, Universit{\'e} de Li{\`e}ge, Quartier Agora, B{\^a}t B5C, All{\'e}e du 6 ao{\^u}t, 19C, B-4000 Li{\`e}ge, Belgium\\
              \email{mahy@astro.ulg.ac.be}
         \and
         Instituut voor Sterrenkunde, KU Leuven, Celestijnenlaan 200D, Bus 2401, 3001 Leuven, Belgium
         \and
         Laboratoire AIM Paris-Saclay, CEA/IRFU - CNRS/INSU - Universit{\'e} Paris Diderot, Service d'Astrophysique, B{\^a}t. 709, CEA-Saclay, 91191, Gif-sur-Yvette Cedex, France 
   }

   \date{Received September 15, 1996; accepted March 16, 1997}

 
  \abstract
   {}
   {The magnetic star HD\,148937 is the only Galactic Of?p star surrounded by a nebula. The structure of this nebula is particularly complex and is composed, from the center out outwards, of a close bipolar ejecta nebula (NGC\,6164/5), an ellipsoidal wind-blown shell, and a spherically symmetric Str{\"o}mgren sphere. The exact formation process of this nebula and its precise relation to the star's evolution remain unknown.}
   {We analyzed infrared {\it Spitzer}  IRS and far-infrared {\it Herschel}/PACS observations of the NGC\,6164/5 nebula. The {\it Herschel} imaging allowed us to constrain the global morphology of the nebula. We also combined the infrared spectra with optical spectra of the central star to constrain its evolutionary status. We used these data to derive the abundances in the ejected material. To relate this information to the evolutionary status of the star, we also determined the fundamental parameters of HD\,148937 using the CMFGEN atmosphere code.}
   {The H$\alpha$ image displays a bipolar or "8"-shaped ionized nebula, whilst the infrared images show dust to be more concentrated around the central object. We determine nebular abundance ratios of $\mathrm{N/O} = 1.06$ close to the star, and $\mathrm{N/O} = 1.54$ in the bright lobe constituting NGC\,6164. Interestingly, the parts of the nebula located further from HD\,148937 appear more enriched in stellar material than the part located closer to the star. Evolutionary tracks suggest that these ejecta have occured $\sim 1.2-1.3$ and $\sim 0.6$\,Myrs ago, respectively. In addition, we derive abundances of argon for the nebula compatible with the solar values and we find a depletion of neon and sulfur. The combined analyses of the known kinematics and of the new abundances of the nebula suggest either a helical morphology for the nebula, possibly linked to the magnetic geometry, or the occurrence of a binary merger.}
   {}


   \maketitle
%
   \section{Introduction}
   \label{sec:Intro}

   With its Of?p spectral type, HD\,148937 belongs to a small class of peculiar stars that has triggered quite some interest in recent years. This category of stars was defined by \citet{wal72} to designate stars displaying some similarities to Of supergiants but also clearly different characteristics - in particular, the presence of strong \ion{C}{iii}~4650\AA\ emission lines. HD\,148937 was amongst the first objects classified as Of?p, but was only recently the target of in-depth studies. 

   HD\,148937 is the brightest and hottest massive star to possess a strong magnetic field \citep{martins12}. The effective temperature of this object reaches $\sim 40000$\,K, its gravity  $\logg \sim 4.0$, and its mass approximately 50--60\,M$_{\odot}$ \citep{naz08a,martins15}. It displays chemical enrichment, but with a level comparable to O supergiants \citep{martins15}. It is a slow rotator, as evidenced by its narrow spectral lines ($\vsini \,\leq 45$\,km\,s$^{-1}$, \citealt{naz10,wade12,martins15}). Its magnetic field was discovered by \citet{hub08} and was subsequently monitored by \citet{wade12} who derived a bipolar field strength of 1kG, confirming the pole-on geometry derived by \citet{naz10}.

The strong magnetic field is able to channel the stellar wind of HD\,148937 towards its (magnetic) equator. As the magnetic axis is tilted with respect to the rotation axis, the dense equatorial regions are seen under different angles during the stellar rotational cycle. This explains why the star presents line profile variations of its Balmer and \ion{He}{i} lines, which are partially formed in this confined wind region. The spectral changes detected for HD\,148937 are, however, of a smaller amplitude than for other Of?p stars \citep{naze08b}. This is explained by a nearly constant pole-on configuration \citep{naz10} that was confirmed later with spectropolarimetric observations \citep{wade12}. In view of their interpretation, these changes should be periodic and they indeed are, with a period of only 7.03~days \citep{naze08b}, whilst the other Galactic Of?p stars display much longer periods (73 days for CPD$-$28$\degr$\,2561, 158 days for NGC1624$-$2, 538 days for HD\,191612, and 55 years for HD108). Finally, HD\,148937 emits an intense X-ray emission which is one order of magnitude brighter than "normal" O-stars but similar to other magnetic O-stars \citep{naz10,naz12,naz14}. The derived location of the emission region ($\sim$1\,R$_{\odot}$ above the photosphere) as well as the X-ray strength are compatible with hot plasma arising from the head-on collision between channeled wind flows from the two hemispheres, even though the softness of the emission, compared to e.g. $\theta^1$\,Ori\,C, remains unexplained \citep{naz12,naz14}.

HD\,148937 is located at the center of NGC\,6164/5, a bipolar nebula that was studied by \citet{leitherer87} and \citet{dufour88}. The latter showed that the composition of the bipolar structure NGC\,6164/5 is chemically peculiar with an overabundance in nitrogen and in helium and a depletion in oxygen and in neon.

The kinematics of the nebula was analyzed in detail in several studies and seems to be also peculiar. \citet{cathpole70} found that the brightest lobes (constituting the brightest parts of NGC\,6164 and NGC\,6165) of the nebula were moving nonisotropically with radial velocities (RVs) of about $+21\,\kms$ and $-43\,\kms$ for the NW and the SE lobes, respectively. These results were refined and confirmed by \citet{pismis74} and later by \citet{carranza86} and \citet{leitherer87}. However, the kinematics of the nebula as a whole is not as simple as this would suggest because the inner parts of the nebula present an independent kinematic structure with different parts approaching and/or receding.

When one looks carefully at the interstellar environment that the star is embedded in, the star and the nebula appear further surrounded by an ellipsoidal stellar-wind-blown shell with a radius of about $12\arcmin$ and, further away, by a Strömgren sphere with a radius of about $1.1\degr$ \citep{leitherer87}.  The origin of such complex motions, as well as the nature of the global morphology, are still open questions.

In the present paper, we investigate the {\it Spitzer} IRS and {\it Herschel}/PACS data of the nebula to further constrain the properties of NGC\,6164/5. We also put these results in perspective through a new modeling of the central star with the CMFGEN atmosphere code \citep{hillier98}. The abundances determined in both environments are compared to trace back the evolution of HD\,148937 and to understand how such a nebula was created. The paper is organized as follows. In Section\,\ref{sec:Obs}, the observations and data reduction procedures are presented. In Section\,\ref{sec:Morpho}, we describe the global morphology of the nebula and a general summary of its kinematics is provided in Section\,\ref{sec:kin}. We analyze the {\it Spitzer} IRS and the {\it Herschel}/PACS spectra obtained at nine different positions and in two different regions across the nebula, respectively, in Section\,\ref{sec:Emission}. We then model the central star with an atmosphere code in Section\,\ref{sec:Modelling}. We discuss the results and constrain the evolution as well as the origin of the nebula in Section\,\ref{sec:Discussion}. Finally, we provide the conclusions in Section\,\ref{sec:Conclusion}.

   \section{Observations and data reduction}
   \label{sec:Obs}
   \subsection{Nebula observations}
   \subsubsection{Infrared data}
   \label{subsec:ir}

   The far-infrared observations were obtained, with the PACS photometer \citep{poglitsch10} onboard the {\it Herschel} spacecraft \citep{pilbratt10}, in the framework of the mass-loss of evolved stars (MESS) guaranteed time key program \citep{groenewegen11}. The PACS imaging observations were carried out on September 2nd, 2011, which corresponds to {\it Herschel}'s observational day (OD) 842. The observing mode was the scan map in which the telescope slews at constant speed (in our case the "medium" speed of $20\arcsec$/s) along parallel lines in order to cover the required area of the sky. Two orthogonal scan maps were obtained for each filter. The observation identification numbers (obsID) of the four scans are 1342227809, 1342227810, 1342227811, and 1342227812, each having a duration of $889$~s.

   We use the {Herschel Interactive Processing Environment} \citep[HIPE,][]{ott10} to perform the data reduction up to level 1. Subsequently, we apply the Scanamorphos software \citep{roussel13} to reduce and combine the data with the same wavelength. The pixel size in the final maps is $2\arcsec$ in the blue channel (70 and 100~\mum) and $3\arcsec$ in the red channel (160~\mum). The  point spread function (PSF) full widths at half maximum (FWHMs) are $5.2\arcsec$, $7.7\arcsec$, and $12.0\arcsec$ at 70\,\mum, at 100\,\mum\ and at 160\,\mum, respectively.

   To complete this imaging data set, we also retrieved the {\it WISE} infrared images \citep{wright10} at 12\,\mum\ and at 22\,\mum\ to better constrain the morphology and the properties of the nebula at shorter wavelengths.

   The {\it Herschel}/PACS spectroscopic data of the nebula have been obtained with the PACS integral-field spectrometer that covers the [52--220]~\mum\ wavelength region. This instrument operates simultaneously in two channels: the blue one ranging from 52 to 98~\mum\ and the red one going from 102 to 220~\mum. The nebula is, however, too large with respect to the spectrometer field of view. Therefore, two on-source spectra have been taken, one at the coordinates $RA=16 \mathrm{h}\,33 \mathrm{m}\,44.5 \mathrm{s}$ and $\delta=-48\degr\,06\arcmin\,27.4\arcsec$ and one at coordinates $RA=16\mathrm{h}\,33\mathrm{m}\,42.7\mathrm{s}$ and $\delta=-48\degr\,04\arcmin\,50.0\arcsec$ (see the upper panel of Fig.\,\ref{fig:nomen}). The former region will be referred to in the present paper as the Herschel-1 region (H1) whilst the latter is reported as the Herschel-2 region (H2). In addition, an off-source spectrum has been taken at the coordinate $RA=16 \mathrm{h}\,30\mathrm{m}\,33.0\mathrm{s}$ and $\delta=-48\degr\, 07\arcmin\, 01.1\arcsec$. These spectra allow us to derive the properties of the nebula, in particular the abundances. For each spectrum, simultaneous imaging of a $47\arcsec \times 47\arcsec$ field of view is obtained, resolved in $5\times 5$ square spatial pixels (i.e., spaxels). The two-dimensional field of view is then rearranged along a $1 \times 25 $ pixel entrance slit for the grating via an image slicer employing reflective optics. The resolving power is between 940 and 5500 depending on the wavelength. The two obsIDs related to the spectrum taken in the H2 region are 1342252090, and 1342252091 whilst the two other obsIDs related to the data taken in the H1 region are 1342252092, and 1342252093. The data were reduced with HIPE version 14, using the interactive pipeline script for unchopped line spectroscopy with calibration set 77. In unchopped mode, the level of the continuum is not guaranteed and has no physical meaning. Therefore, we subtract the continuum to only represent the emission lines.

   Nine infrared spectra of HD\,148937 (AORkey 18380032, PI: Morris) and of its nebula NGC\,6164/5 (AORkey 18380800, PI: Morris) were obtained with the Infrared Spectrograph \citep[IRS,][]{houck05} onboard the {\it Spitzer} Space Telescope \citep{werner05}. The observation of the nebula (AORkey 18380800, \#1 to \#8 in the upper panel of Fig.\,\ref{fig:nomen}) was performed in "cluster" mode in which spectra are taken at eight pointed positions whilst the observation of HD\,148937 (AORkey 18380032, \#9 in the upper panel of Fig.\,\ref{fig:nomen}) was performed in "single pointing" mode. The reduced spectra were downloaded from the Cornell Atlas of {\it Spitzer}/IRS Sources (CASSIS, \citealt{lebouteiller11,lebouteiller15}). We use the high-resolution ($R=\lambda/\delta\lambda=600$) spectra observed with the Short-High (SH) and Long-High (LH) modules, covering spectral ranges from 9.9 to 19.6\,\mum\ (SH) and from 18.7 to 37.2\,\mum\ (LH), respectively. The aperture sizes are $4.7\arcsec \times 11.3\arcsec$ and $11.1\arcsec \times 22.3\arcsec$ for the SH and LH spectra, respectively. The spectral extraction uses either a full aperture extraction for which the flux is simply integrated within the aperture or an optimal extraction for which the flux is found by scaling the point-spread function to the source's spatial profile. While the former method is well suited for extended sources, the latter method is only reliable for point-like sources. For this reason, for the eight spectra of the nebula, we used the full aperture extraction, as the emission is always extended in our observations and the optimal extraction for the spectrum taken at the position of HD\,148937 (\# 9 in the upper panel of Fig.\,\ref{fig:nomen}). Unfortunately, we were not able to analyze the LH spectrum in the optimal extraction mode because the background emission was too strong to allow its extraction. We therefore only use the SH spectrum at that position. For the spectra across the nebula, a scaling factor has been applied to the SH spectra in order to align it with the continuum of the LH spectrum. This scaling factor is requested to quantify the correction for aperture losses and is reported for each spectrum in Table\,\ref{tab:spitzer}. 

   The locations at which the different infrared spectra have been taken are displayed in the upper panel of Fig.\,\ref{fig:nomen}. As a comparison, and because we use it later in the present paper, we also show the areas of the nebula analyzed by \citet{leitherer87} and \citet{dufour88} in the lower panel of Fig.\,\ref{fig:nomen}.
   \subsubsection{Optical data}
   \label{subsubsec:gemini}

   Optical images of NGC\,6164/5 have been retrieved from the Gemini Science Archives. A series of $120 s$ exposures were obtained with GMOS-S imager in three different filters: [\ion{S}{ii}] ($\lambda_c = 6720\AA$, FWHM~=~44.4\AA), [\ion{O}{iii}] ($\lambda_c = 4990\AA$, FWHM~=~44.3\AA) and H$\alpha + [\ion{N}{ii}]$ ($\lambda_c = 6560\AA$, FWHM~=~71.6\AA). Unfortunately, no continuum image was taken along with those nebular images. The standard reduction has been performed.

   To calibrate the flux of the H$\alpha$ image of the nebula, we used the {\it HST} image in the H$\alpha$ filter obtained in 1991 with the WFPC equipped with the F656N filter. This filter is centered on H$\alpha$ but the neighbouring [NII] lines also contribute to the recorded emission (to the level of 30\%). Reduced {\it HST} observations were downloaded from the archives. They focused only on the southeast lobe of the nebula (NGC\,6165) and were reported by \citet{sco95}. These data were combined to make a mosaic using the task {\it wmosaic} under IRAF and the flux calibration is obtained by multiplying the recorded ADUs by the PHOTFLAM keyword and the filter width ($19~\AA$), by dividing by the exposure time ($2 \times 800$s) and by correcting for 30\% of the [\ion{N}{ii}] contribution. 

   \subsection{Stellar observations}
   
   To determine the stellar parameters and the abundances of HD\,148937, we also retrieved the spectra of HD\,148937 taken with FEROS and ESPaDOnS presented by \citet{naze08b} and \citet{wade12}, respectively. The analysis of the variations observed in the spectra has been done in both papers and will not be repeated in the present study. The reduction processes of both sets of data were the same as explained in the two papers.
   
   \section{Morphology of the nebula}
   \label{sec:Morpho}

\begin{figure*}
   \centering
   \includegraphics[width=5cm,bb=306 159 531 384,clip]{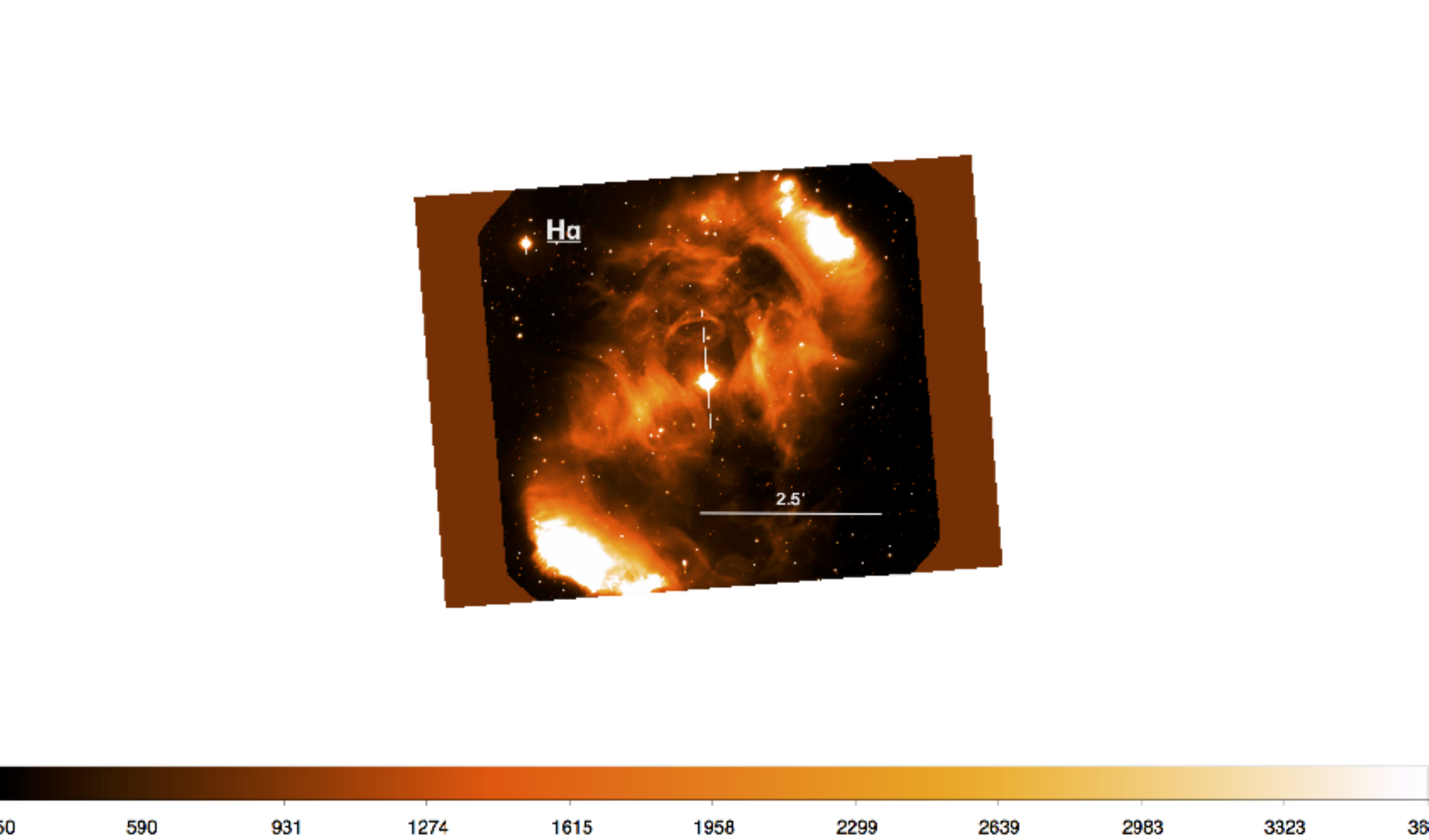}
   \includegraphics[width=5cm,bb=220 60 636 476,clip]{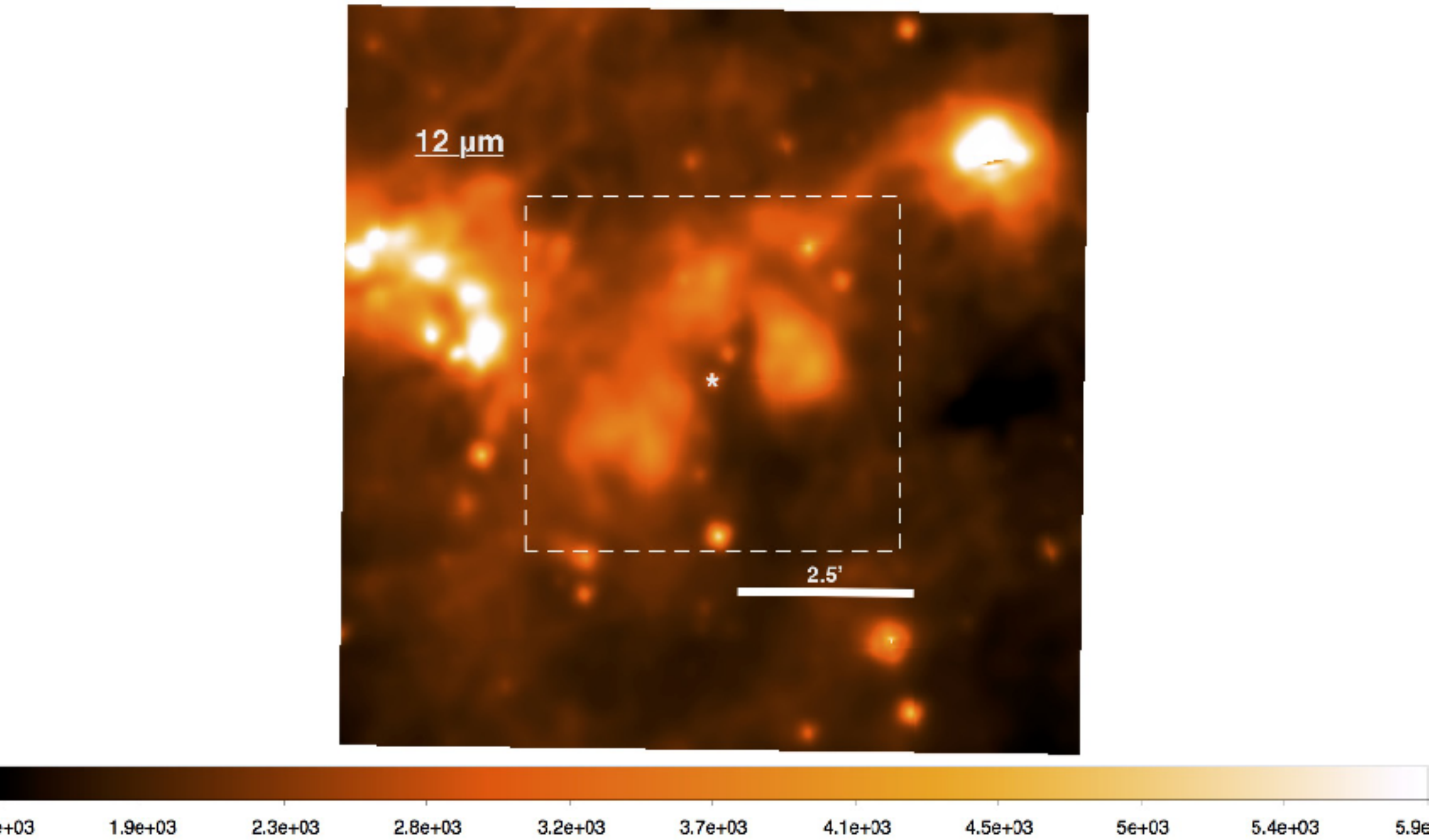}
   \includegraphics[width=5cm,bb=220 60 636 476,clip]{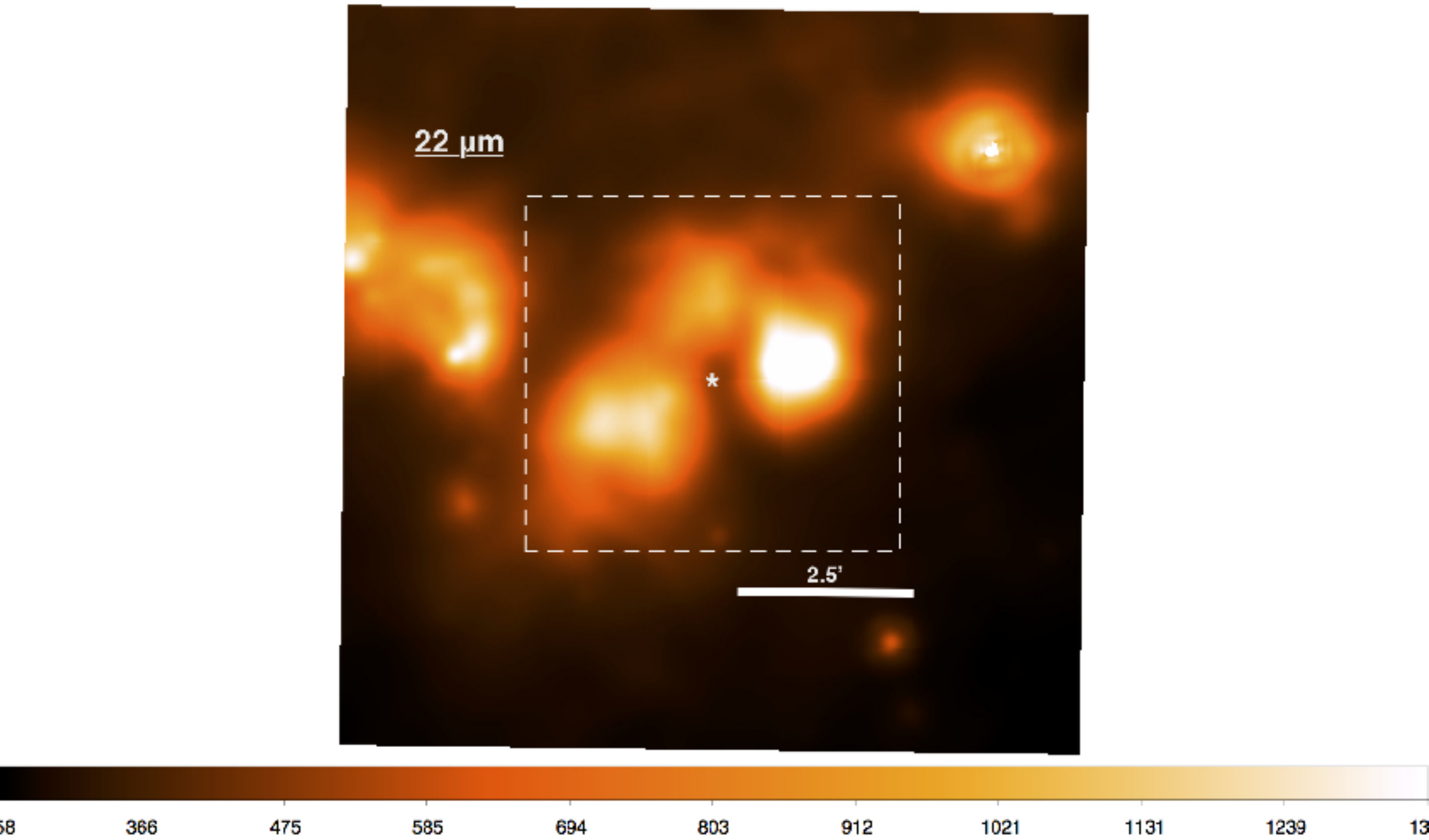}
   \includegraphics[width=5cm,bb=220 60 636 476,clip]{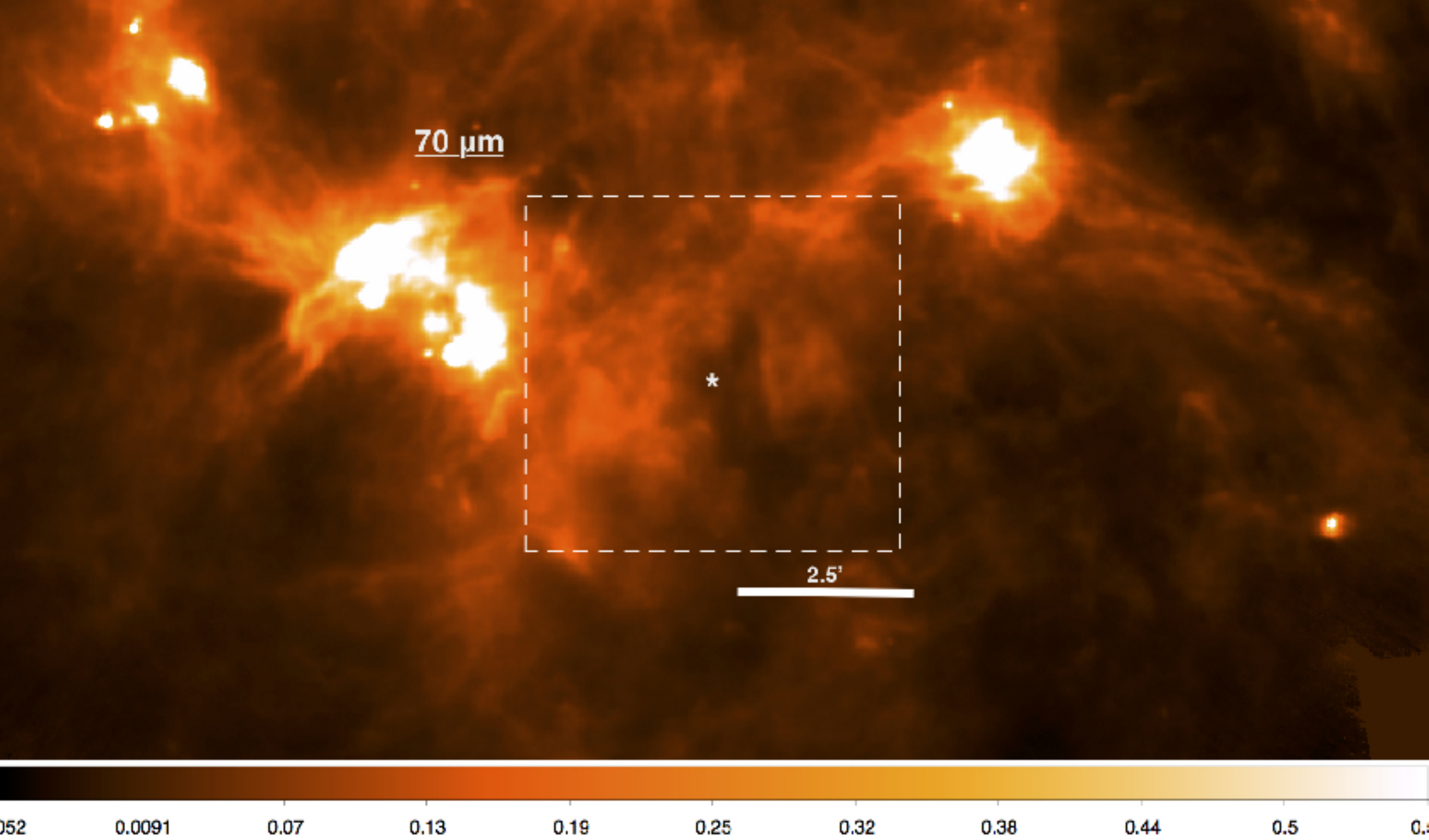}
   \includegraphics[width=5cm,bb=220 60 636 476,clip]{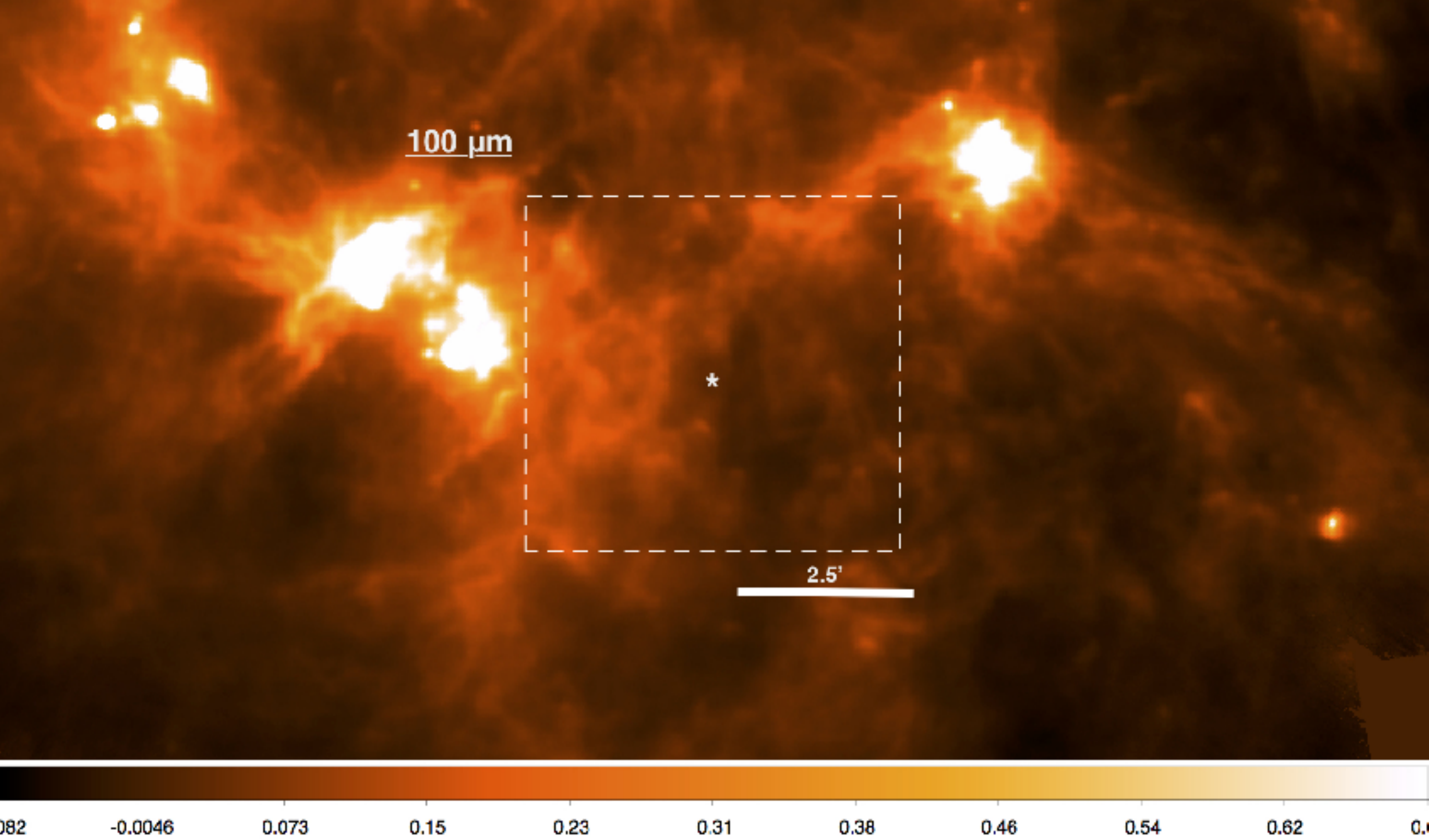}
   \includegraphics[width=5cm,bb=220 60 636 476,clip]{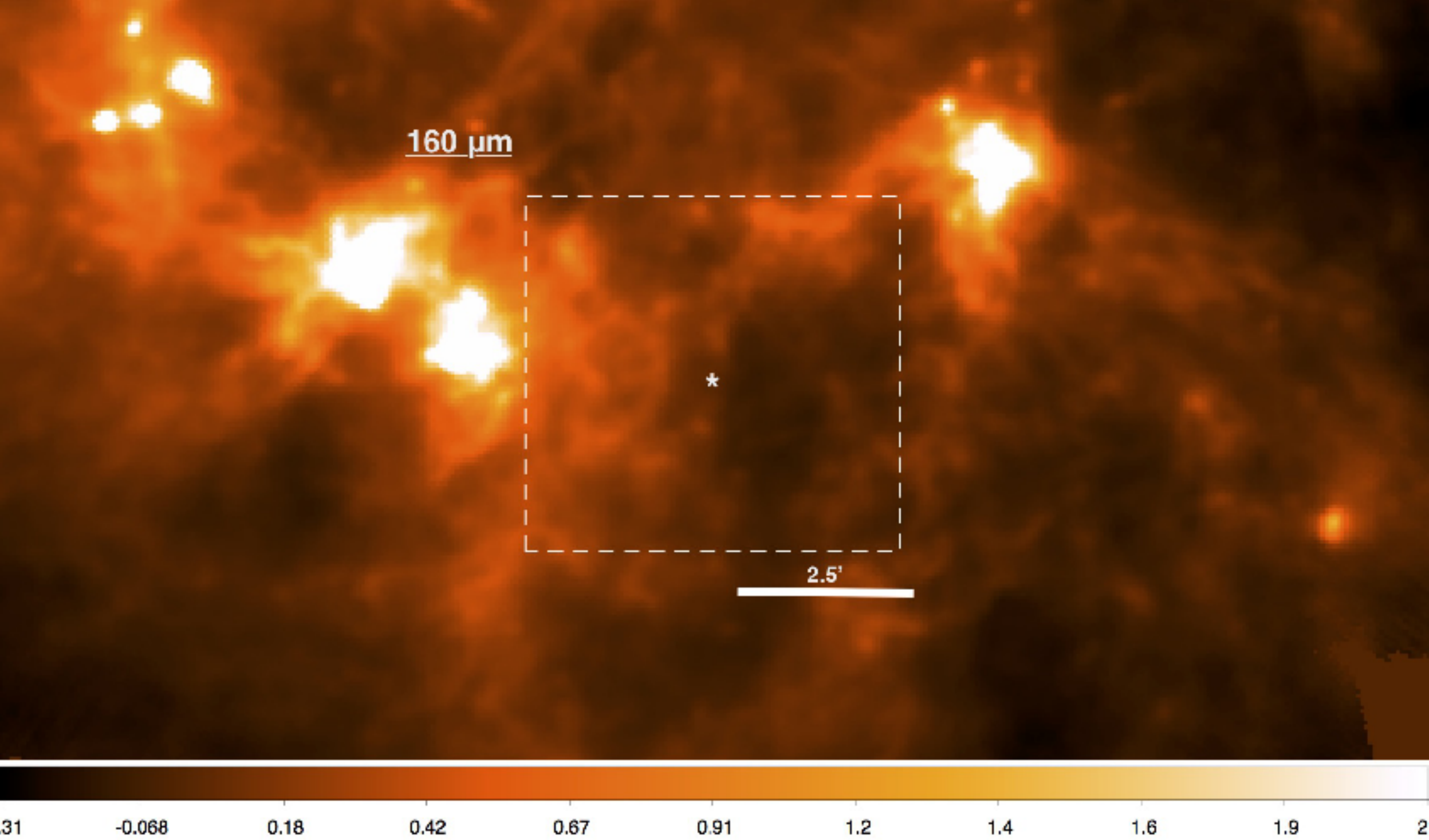}
   \caption{Images of NGC6164/5 around HD\,148937. {\it Top:} the Gemini H$\alpha + [\ion{N}{ii}]$ image ({\it left}), the { \it WISE}~12~\mum\ and 22~\mum\ images ({\it middle} and {\it right}, respectively). {\it Bottom:} PACS images of the nebula at 70~\mum, 100~\mum\ and 160~\mum\ from {\it left} to {\it right}, respectively. The white asterisk indicates the position of HD\,148937. The dash-lined box provides the size of the H$\alpha$ image because the sizes of these images are different. North is up and east is left. }\label{fig:images}%
   \end{figure*}

The H$\alpha$ and the infrared images of HD\,148937 and of its surrounding nebula NGC\,6164/5 at 12\,\mum, 22\,\mum,  70\,\mum, 100\,\mum\ and 160\,\mum\ are shown in Fig.~\ref{fig:images}.

   \begin{figure}
   \centering
   \includegraphics[width=7cm,bb=90 78 499 435,clip]{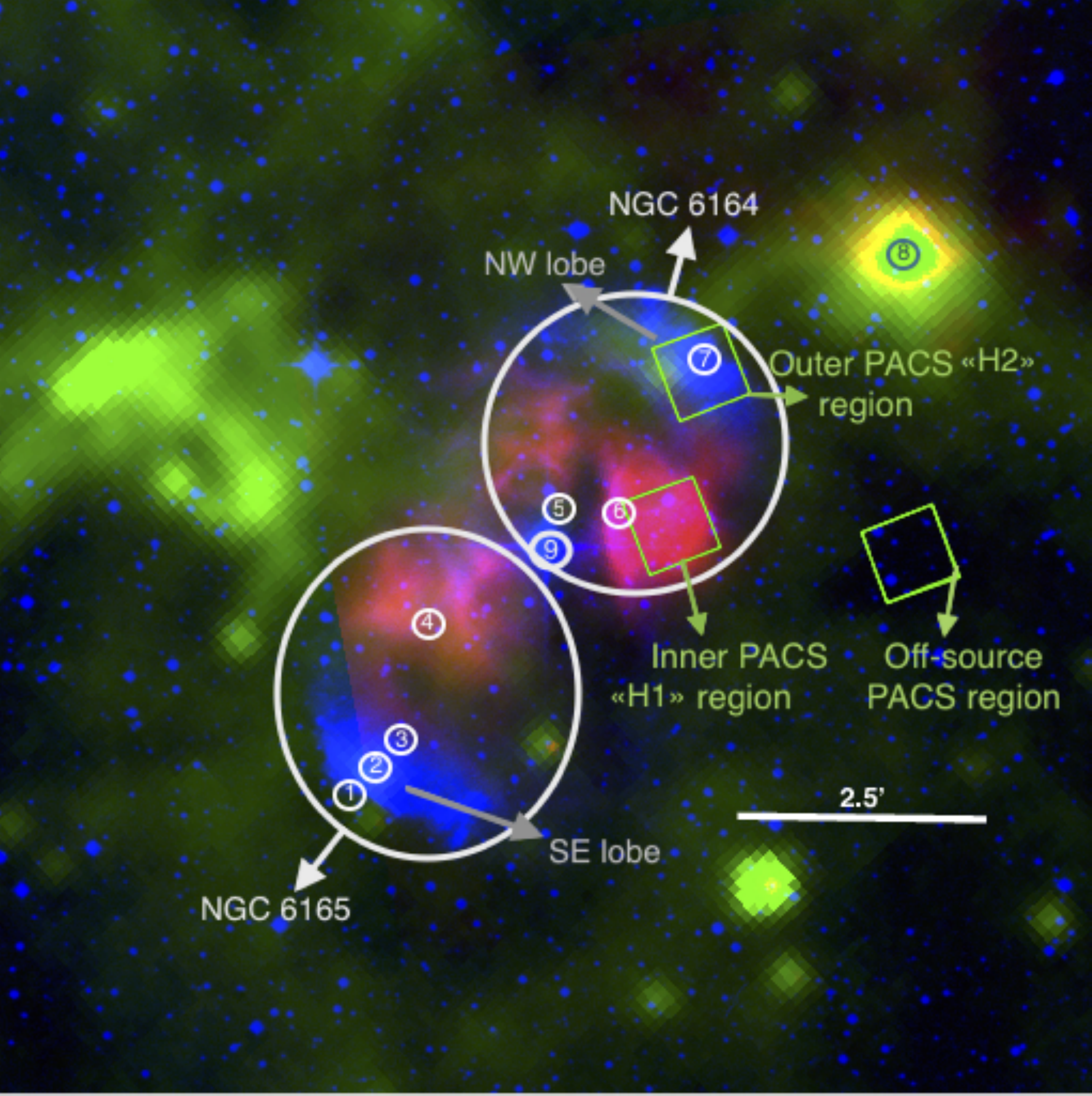}
   \includegraphics[width=7cm,bb=324 191 739 560,clip]{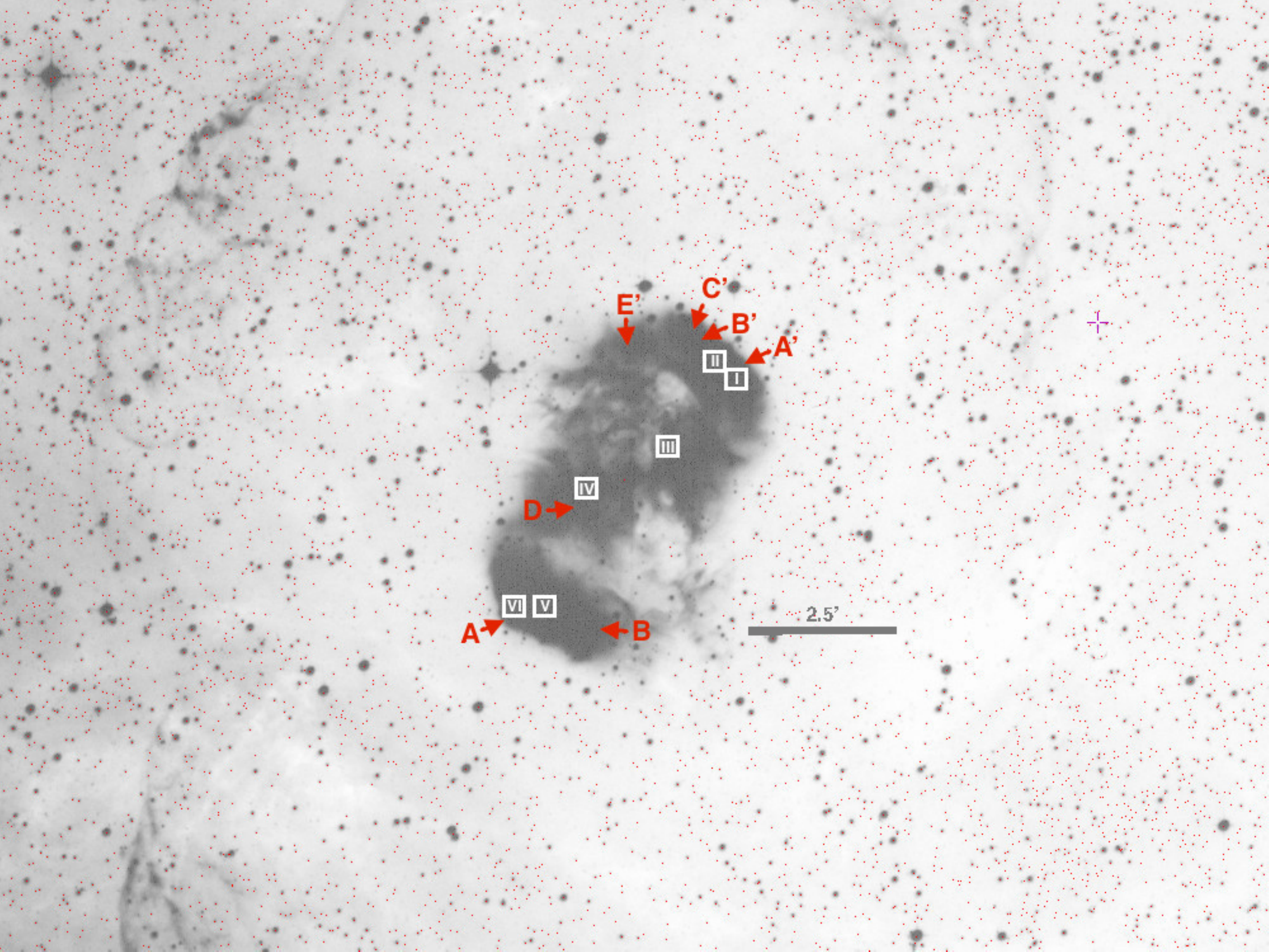}
   \caption{{\it Top:} General picture of the nebula NGC\,6164/5 and definition of the nomenclature used in the present paper. North is up and east is left. The small white circles represent the positions of the {\it Spitzer} IRS spectra whilst the green squares give the footprints of the {\it Herschel}/PACS field of view. The NW and SE lobes are indicated by gray arrows and the inner zones of ejecta are represented in red. {\it Bottom:} H$\alpha$ image of NGC\,6164/5 taken from the SuperCosmos H$\alpha$ Survey \citep{parker05}. The white boxes represent the areas of the nebula analyzed by \citet{leitherer87} whilst the red arrows represent the positions analyzed by \citet{dufour88}. }\label{fig:nomen}%
   \end{figure}

The H$\alpha$ image reveals an ionized nebula, oriented to the NW-SE axis, with a bipolar or "8" shape centered on the Of?p HD\,148937. This ionized nebula is composed of different visible structures:  two brighter lobes located at SE and NW edges of the nebula (in blue and indicated by gray arrows in the upper panel of Fig.\,\ref{fig:nomen}) and three fainter zones of ejecta located closer to the central star (in red in the upper panel of Fig.\,\ref{fig:nomen}). The whole nebula is formed by NGC\,6164 at NW and NGC\,6165 at SE (see the large white circles in Fig.\,\ref{fig:nomen}). In the ionized nebula, the maxima of emission in these two regions are located at about $150\arcsec$ and $185\arcsec$ from HD\,148937 in the NW and the SE directions, respectively. Assuming that the distance between the star and Earth is about 1.3\,kpc (the distance of Ara\,OB1, \citealt{herbst77}, which HD\,148937 is assumed to be embedded in), the projected separation would then be equal to 0.95\,pc and 1.16\,pc for NGC\,6164 and NGC\,6165, respectively. \citet{leitherer87} estimated the inclination between the plane in which the nebula has been ejected and the line of sight to $85\degr$ even though, according to these authors, the inclination can be assumed to be between $60\degr$ and $90\degr$.

   \begin{figure*}
   \centering
   \includegraphics[width=13cm,bb=205 49 639 496,clip]{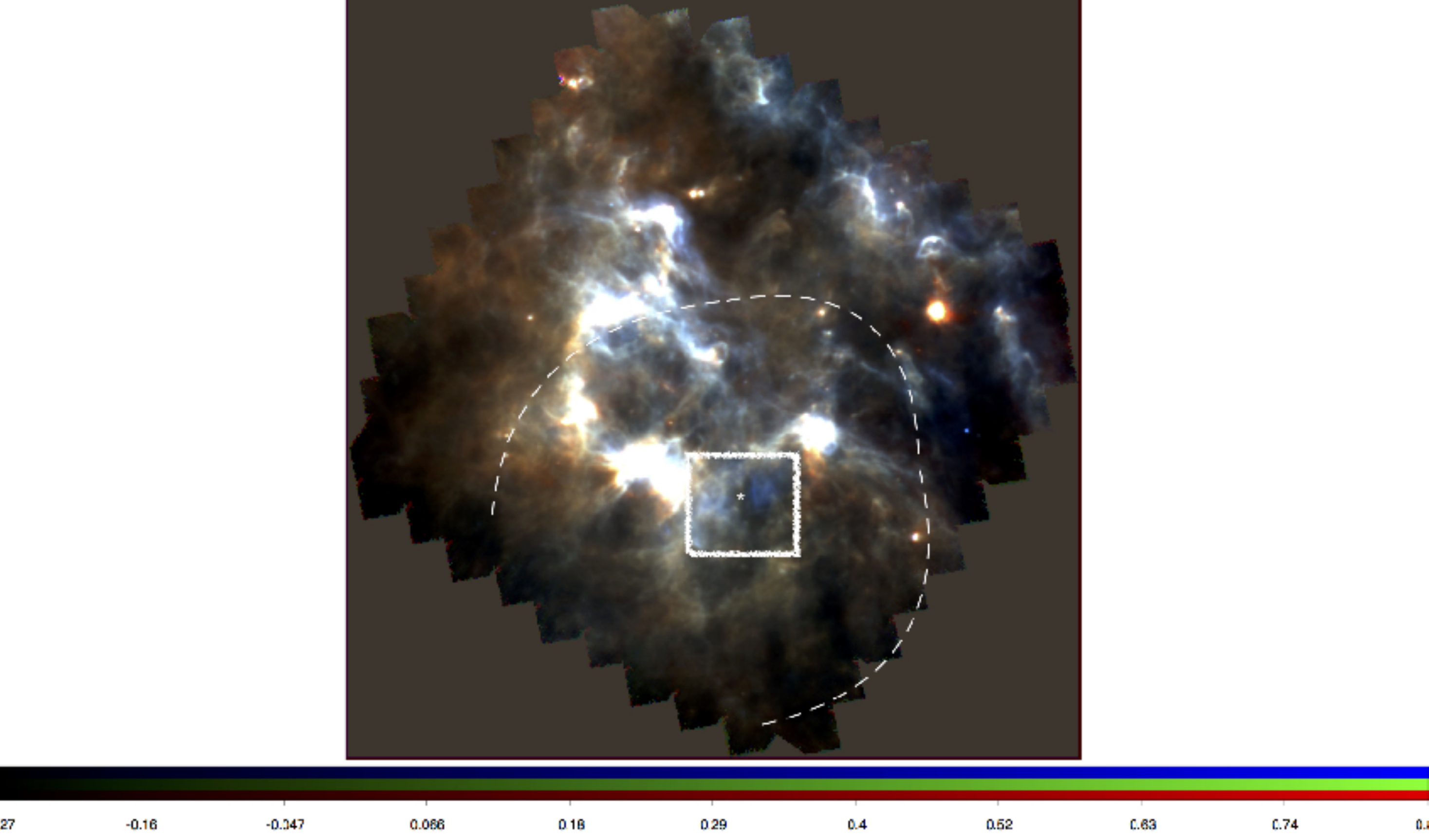}
   \caption{Three-color image of PACS photometer field of view with 70\,\mum\ in blue, 100\,\mum\ in green and 160\,\mum\ in red. North is up and east is left. The box indicates the location of the H$\alpha$ image (see Fig.\,\ref{fig:images}). The dashed line represents the location of the wind-blown structure at $12\arcmin$ from HD\,148937 (seen on the H$\alpha$ image). }\label{fig:color}%
   \end{figure*}

In the infrared images, the lobes become fainter whilst the three inner zones of ejecta become brighter until the nebular emission is not distinguishable from the external environment at wavelengths longer than 70\,\mum. This observation suggests that the dust emission is weaker in the lobes although the presence of cooler dust in the lobe cannot be excluded. Indeed, at the inner edges of the lobes, one might detect, from the H$\alpha$ image, a zone of collision. According to \citet{vanmarle11}, only large grains can pass through this zone. By combining a larger distance which the lobes are located at and a possible larger size of dust grains in the lobes, we expect lower dust emission in the lobes at wavelengths shorter than 70\,\mum, as observed. In the 22\,\mum\ image, we see that the zone of ejecta (in red in Fig.\,\ref{fig:nomen}) located at west is particularly bright in comparison to the others. This zone is in fact contaminated by the presence of a dark cloud seen in projection \citep{peretto09}, that makes the analysis of the NGC\,6164/5 more complex. Because of this dark cloud and of the high background emission, we cannot measure accurate flux densities from NGC\,6164/5 only and thus determine an accurate value of the mass of the dust in the nebula NGC\,6164/5. Finally, it is worth noting that the infrared images also show a cavity centered on HD\,148937. This lack of emitting material has probably been formed through the action of the stellar wind. The shape and the size of this cavity are different from one infrared image to another: it has a horseshoe shape in the 12\,\mum, 22\,\mum, and 70\,\mum\ images but appears larger and with an unconstrained shape in the images at longer wavelengths. We can roughly estimate lower limits on the angular distances of $35\arcsec$, $51\arcsec$, and $64\arcsec$, giving distances of 0.22, 0.32 and 0.40\,pc between the star and the eastern, western and northern inner zone of ejecta, respectively, assuming a distance between the star and the Earth of 1.3\,kpc. 

 In Fig.~\ref{fig:color}, we show a larger-scale view using a three-color image based on the PACS data (70~\mum\ in blue, 100~\mum\ in green and 160~\mum\ in red). The nebula around HD\,148937 is barely detectable at the bottom of this image (white box in Fig.\,\ref{fig:color}) and does not constitute the brightest part of the image. Given the position of the nebula and the size of the observed field, it is not possible from this image to detect the Stromgren sphere at $1.1\degr$ observed by \citet{leitherer87} whilst the wind-blown structure at $12\arcmin$ from HD\,148937 is partially in the field of view of PACS (dashed line in Fig.\,\ref{fig:color}), but cannot be distinctly observed because of the background emission.
   
 \section{Kinematics of the nebula}
 \label{sec:kin}

 The kinematics of the whole ionized H$\alpha$ nebula was analyzed by \citet{cathpole70}, \citet{pismis74}, \citet{carranza86} and \citet{leitherer87}. These different authors measured the RVs of the ejecta in four different regions: the NW and SE lobes as well as in the western and the eastern bright parts. Their measurements were, in general, in good agreement. Before comparing the RVs of these different parts, it is worth noting that the velocity of the center-of-mass of HD\,148937 was measured to be $-26$\,\kms\ \citep{naz08a}. A part of the NW lobe (constituting NGC\,6164) is moving away from the observer with RVs between $+21$ and $+55$\,\kms\ but another part of this region is also approaching the observer with a RV of $-3$\,\kms. The SE lobe (forming NGC\,6165) is moving toward the observer with RVs between $-30$ and $-145$\,\kms. The kinematics of the ejecta located between both lobes is more complex. \citet{leitherer87} reported H$\alpha$ profiles reproduced with four Gaussian profiles, giving four different RVs for the ejecta located at the western part of the nebula (RVs~$ =-70$, $-19$, $+33$, and $+63$\,\kms\ were reported) whilst they used only two Gaussian profiles to fit the H$\alpha$ profiles observed in the eastern part of the nebula, giving RVs of $-79$ and $-13$\,\kms. This complex kinematics suggests that the bipolar or "8"-shaped nebula displayed in the H$\alpha$ image could thus be the result of the projection of a helical structure centered on HD\,148937, as explained in further detail by \citet{carranza86}. 

   \section{Emission line spectrum}
   \label{sec:Emission}

   \subsection{{\it Spitzer} IRS spectra}

\begin{figure*}
\centering 
\includegraphics[width=12cm,bb=37 7 524 395,clip]{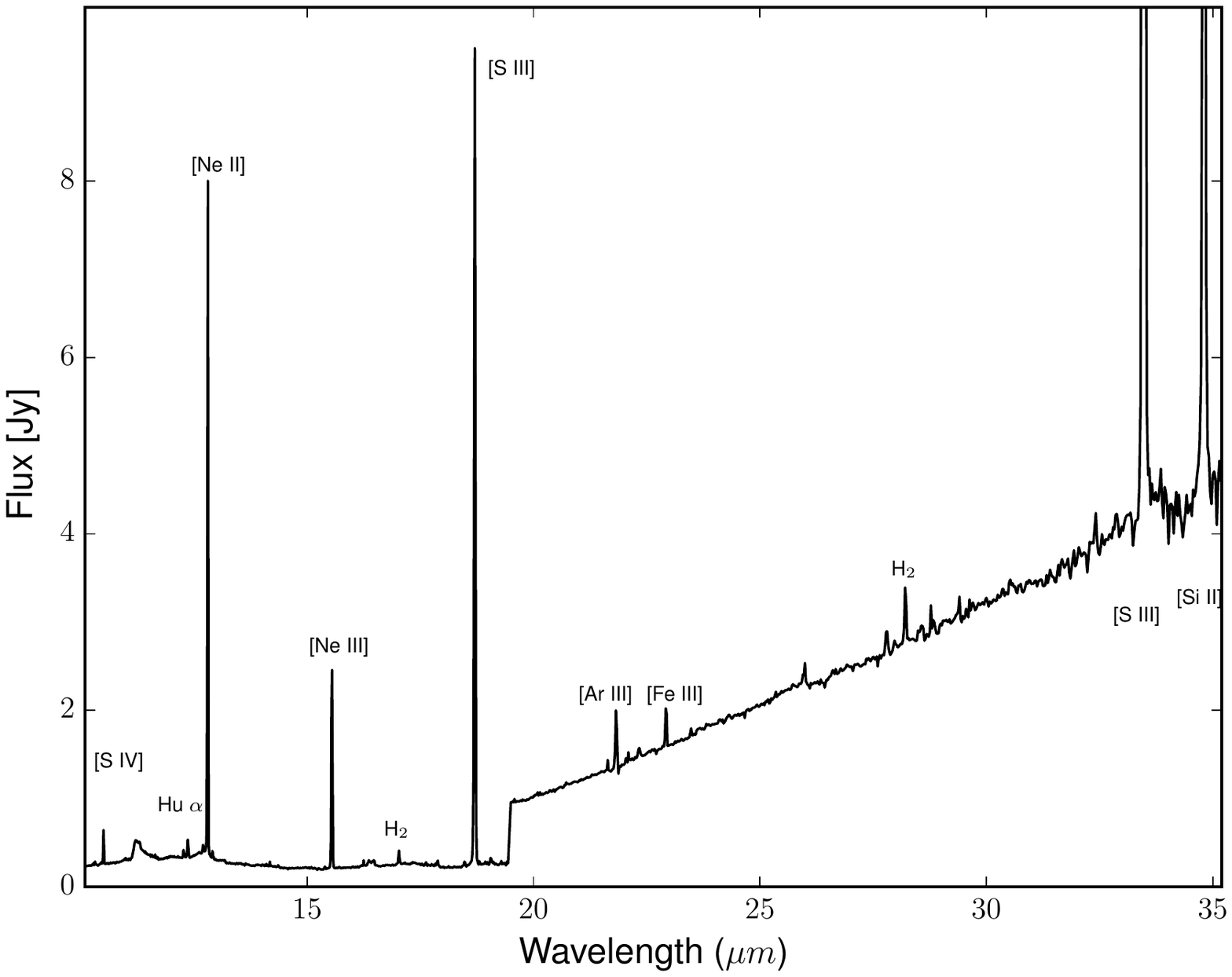}
\caption{{\it Spitzer} IRS spectrum taken at position \#2 (see Fig.\,\ref{fig:nomen}). The step at 19\,\mum\ is due to the different apertures used for the SH and LH spectra. }\label{fig:spitzerspec}
\end{figure*}

   Eight mid-infrared spectra have been taken at different positions across the nebula and one on HD\,148937. As previously mentioned, the exact locations of these nine spectra are displayed in Fig.\,\ref{fig:nomen}. These data show spectral lines that can provide important constraints on the physical conditions in the nebula. As an example, we show in Fig\,\ref{fig:spitzerspec} the merged {\it Spitzer} IRS spectrum (SH$+$LH) taken at position \#2 (see the upper panel of Fig.\,\ref{fig:nomen}) and where the SH spectrum is not corrected for the scaling factor. In addition, because HD\,148937 is a point-like source, its LH spectrum was extracted under the optimal mode. The latter is, however, not usable because the source is highly contaminated by the background level. Therefore, the measurements for HD\,148937 are taken only on the SH spectrum. 

   We measure the emission line flux densities by fitting a Gaussian profile to the main line profiles. The following emission lines are detected in the different spectra: [\ion{S}{iv}]\,10.5\,\mum, Hu\,$\alpha$\,12.4\,\mum, [\ion{Ne}{ii}]\,12.8\,\mum, [\ion{Ne}{iii}]\,15.5\,\mum, [\ion{S}{iii}]\,18.7\,\mum, [\ion{Ar}{iii}]\,21.8\,\mum, [\ion{Fe}{iii}]\,22.9\,\mum, H$_2$\,28.2\,\mum, [\ion{S}{iii}]\,33.5\,\mum\ and [\ion{Si}{ii}]\,34.8\,\mum\ as well as PAH features at 11.3\,\mum. The correction factors as well as the corrected fluxes measured for the different lines are listed in Table\,\ref{tab:spitzer}. We also emphasize that the {\it Spitzer} position \#8 is located outside the nebula, on an infrared source. The measured fluxes are thus contaminated by this external source and are thus not representative of the fluxes measured across the nebula.

\begin{table*}
\caption{Line fluxes measured on the {\it Spitzer} IRS spectra.}             
\label{tab:spitzer}      
\centering                          
\begin{tabular}{l c c c c c c c c c c}        
\hline\hline                 
Line & Wavelength & \multicolumn{9}{c}{Corrected flux}  \\    
 &  & \multicolumn{9}{c}{$\times 10^{-15}$}   \\
 &  [\mum] &  \multicolumn{9}{c}{[W m$^{-2}$/aperture]} \\
\hline                      
&           &     1    &  2 & 3 & 4 & 5 & 6 & 7 & 8 & 9    \\
\hline         
\protect[\ion{S}{iv}]$^{a,c}$    & 10.5 & 0.21  & 0.72  & 0.96 & 1.69 & 4.04 & 0.74 & 0.61 & 3.06  & 0.04\\
Hu\,$\alpha$ $^{b,c}$            & 12.4 & 0.21  & 0.42  & 0.18 & 0.20 & 0.18 & 0.14 & 0.25 & 1.96  & 0.03\\
\protect[\ion{Ne}{ii}]$^{a,c}$   & 12.8 & 4.86  & 11.03 & 4.58 & 3.12 & 3.10 & 3.89 & 6.83 & 21.02 & 0.26\\
\protect[\ion{Ne}{iii}]$^{a,c}$  & 15.5 & 0.73  & 3.08  & 2.89 & 2.33 & 3.97 & 1.07 & 2.13 & 12.63 & 0.09\\
\protect[\ion{S}{iii}]$^{a,c}$   & 18.7 & 3.66  & 9.30  & 4.15 & 2.63 & 3.42 & 2.13 & 5.12 & 35.35 & 0.11\\
\protect[\ion{Ar}{iii}]$^{b}$    & 21.8 & 0.06  & 0.16  & 0.11 & 0.08 & 0.13 & 0.05 & 0.13 & 1.15  & --  \\
\protect[\ion{Fe}{iii}]$^{b}$    & 22.9 & 0.03  & 0.11  & 0.05 & 0.09 & 0.04 & 0.03 & 0.04 & 0.60  & --  \\
H$_2$ $^{b}$                     & 28.2 & 0.12  & 0.11  & 0.12 & 0.13 & 0.10 & 0.13 & 0.16 &  --   &  -- \\
\protect[\ion{S}{iii}]$^{a}$     & 33.5 & 6.06  & 7.69  & 6.45 & 4.76 & 5.22 & 3.77 & 6.72 & 37.06 &  -- \\
\protect[\ion{Si}{ii}]$^{a}$     & 34.8 & 2.72  & 3.00  & 2.98 & 3.63 & 3.20 & 3.59 & 3.25 & 10.61 &  -- \\
\hline                                   
\end{tabular}
\tablefoot{ $^{a}$ Flux uncertainties $\leq$ 10\%. \\   $^{b}$ Flux uncertainties between 10\% and 20\%. \\   $^{c}$ The measured fluxes are corrected by the following scale factors: 4.7, 3.8, 4.7, 4.8, 4.3, 5.6, 4.6, 3.0 for spectra 1 to 8, respectively.}
\end{table*}
   
\subsubsection{Electron density}
\label{elec}

The ratio between the [\ion{S}{iii}]\,33.5\,\mum\ and [\ion{S}{iii}]\,18.7\,\mum\ lines can be used to derive the electron density in the ionized nebula surrounding HD148937. 

\begin{table*}
\caption{Electron temperatures, electron densities and abundances across the nebula.}             
\label{tab:abun}      
\centering                          
\begin{tabular}{l c c c c c c}        
\hline\hline                 
Regions  & $T_e$      & $n_e$       & N/O   &  Ar/H  &  Ne/H  & S/H \\
         & [K]        & [cm$^{-3}$] &       & \multicolumn{3}{}{[$12 + \log X/H$]}\\
\hline
1        & $7000\pm 2500$  & $ 80 \pm 50$    & --   & $>5.8$ & 7.7  & $>6.5$  \\
2        & $7000\pm 2500$  & $930 \pm 70$    & --   & $>6.0$ & 7.8  & $>6.6$  \\
3        & $7000\pm 2500$  & $120 \pm 60$    & --   & $>6.1$ & 7.8  & $>6.6$  \\
4        & $7000\pm 2500$  & $ 50 \pm 40$    & --   & $>6.0$ & 7.6  & $>6.4$  \\
5        & $7000\pm 2500$  & $135 \pm 60$    & --   & $>6.2$ & 7.7  & $>6.5$  \\
6        & $7000\pm 2500$  & $ 70 \pm 50$    & --   & $>5.9$ & 7.8  & $>6.4$  \\
7        & $7000\pm 2500$  & $300 \pm 70$    & --   & $>6.1$ & 7.8  & $>6.6$  \\
8        & $7000\pm 2500$  & $600 \pm 30$    & --   & $>6.1$ & 7.4  & $>6.5$  \\
9        & $7000\pm 2500$  & --              & --   & $>6.1$ & 8.3  & $>5.5$  \\
H1       & $7000\pm 2500$  & $ 65 \pm 25$    & 1.06 &  --   & --   & --   \\
H2       & $7000\pm 2500$  & $125 \pm 15$    & 1.54 &  --   & --   & --   \\
A$^{a}$  & $6400 \pm 600$  & $1000 \pm 200$  & 1.15 & $>6.3$& --   & 7.1  \\
B$^{a}$  & $7400 \pm 400$  & $1400 \pm 400$  & 0.78 & $>6.3$& --   &$>6.3$\\
D$^{a}$  & $(7500)$        & $220 \pm 100$   & 0.19 & $>6.3$& --   &$>6.0$\\
A'$^{a}$ & $7600 \pm 500$  & $1200 \pm 600$  & 1.55 & $>6.0$& 7.5  &$>6.2$\\
B'$^{a}$ & $8800 \pm 500$  & $350 \pm 200 $  & 1.51 & $>6.2$& 7.3  &$>5.7$\\
C'$^{a}$ & $8500 \pm 500$  & $630 \pm 300 $  & 1.23 & $>6.1$& --   &$>5.7$\\
E'$^{a}$ & $8300 \pm 700$  & $7000 \pm 3000$ & 1.10 & $>6.0$& 7.3  &$>6.3$\\
I$^{b}$  & --              & --              & --   &  --   & --   & --   \\
II$^{b}$ & $6800 \pm 800$  & $10000 \pm 5000$& --   &  --   & --   & --   \\
III$^{b}$& $<16600$        & --              & --   &  --   & --   & --   \\
IV$^{b}$ & $<21400$        &  $500 \pm 350$  & --   &  --   & --   & --   \\
V$^{b}$  & $6600 \pm 800$  &  $6300 \pm 3700$& --   &  --   & --   & --   \\
VI$^{b}$ & --              & --              & --   &  --   & --   & --   \\
\hline
\end{tabular}
\tablefoot{$^{a}$: Values from \citet{dufour88}\\  $^{b}$: Values from \citet{leitherer87}\\ } 
\end{table*}

To determine the emission line intensities of the [\ion{S}{iii}]\,18.7\,\mum\ line, we must apply a correction factor to take the aperture losses into account. This correction factor is estimated by aligning the continuum flux of the SH spectra to that of the LH spectra in the small overlap between the two spectra, except for the region near the star (\#9) where this correction is impossible (see Sect.\,\ref{subsec:ir}). In addition to the error on the ratio between the intensities of the [\ion{S}{iii}] lines ($\sim 5\%$), the global error on the determination of the electron density is dominated by the errors on the correction factor and on the electron temperature. 

To measure the electron densities in the different regions, we use the nebular/IRAF package. To this purpose, we need an estimate of the electron temperature across the nebula. \citet{leitherer87} as well as \citet{dufour88} investigated the electron temperature at different locations (not necessarily the same positions as the {\bf \it Spitzer} IRS spectra, see Fig.\,\ref{fig:nomen}) in the nebula. Even though the electron temperatures close to the lobes are in agreement between these two studies, the other values obtained in the other parts of the nebula differ from one analysis to another. Furthermore, \citet{leitherer87} provided two upper limits for the electron temperature. We try to use these upper limits to determine the electron densities in our spectra but in some cases, these upper limits are too high to derive a physical value for the electron density from the observed line ratios. Therefore, we consider in this section as well as for our entire analysis a global electron temperature of 7000\,K (compatible with values reported by these authors). We also assume a conservative uncertainty of 2500\,K on the electron temperature.

From this value, we obtain very small electron densities across the nebula except for the lobes where we find $n_e = 300 \pm 70$\,cm$^{-3}$ and $n_e = 930 \pm 70$\,cm$^{-3}$ in the central parts of the NW and the SE lobes, respectively. Elsewhere, at the different locations of the {\it Spitzer} IRS spectra, the electron densities are closer to 100\,cm$^{-3}$. At position \#4, the electron density is relatively low, at the limit of measurability. We note that, if we decrease $T_e$ to 5000\,K, the electron density in this area becomes $70 \pm 20$\,cm$^{-3}$. We emphasize that we have determined the r.m.s. densities on the H$\alpha$ image at the same positions where the {\it Spitzer} data were taken to test our estimates on the electron density. These r.m.s. densities can be taken as lower limits on the electron density but are dependent on the volume that we consider to measure them. As first approximation, we have assumed spherical and cylindrical volumes with radii equal to the aperture sizes of {\it Spitzer}. We note a global agreement between the r.m.s. densities and the electron densities, thus validating our estimations. The derived values of the electron densities are given in Table\,\ref{tab:abun}.

\subsubsection{Abundances}

The {\it Spitzer} IRS spectra do not provide information on CNO abundances across the nebula. They do, however, provide abundances for elements such as sulfur, argon, or neon. To take homogeneous measurements of the hydrogen fluxes, we focus on the flux estimated from the Hu~$\alpha$ hydrogen line at 12.37\,\mum. We then assume a case B recombination with $T_e = 7000$\,K. Table\,\ref{tab:abun} contains the abundances that we derive for the nine {\it Spitzer} IRS spectra. 

To derive the abundances, we use the nebular/IRAF package. The total elemental abundances are difficult to derive because certain elements do not have measurable lines in every ionization state in the {\it Spitzer} wavelength domain. Therefore, we consider that the ionic abundances such as Ar$^{++}$/H$^{+}$, or S$^{++}$/H$^{+}$ are lower limits to the real abundance values. We nevertheless obtain abundances for argon, neon and sulfur compatible with those provided by \citet{dufour88}. Furthermore, the abundances of these three elements are relatively constant across the nebula, with Ar/H compatible with the solar abundance whilst Ne/H and S/H could be considered as depleted in both the inner and outer regions of the nebula.

\subsection{PACS spectra}

\begin{figure}
\centering 
\includegraphics[width=9cm,bb=5 0 537 408,clip]{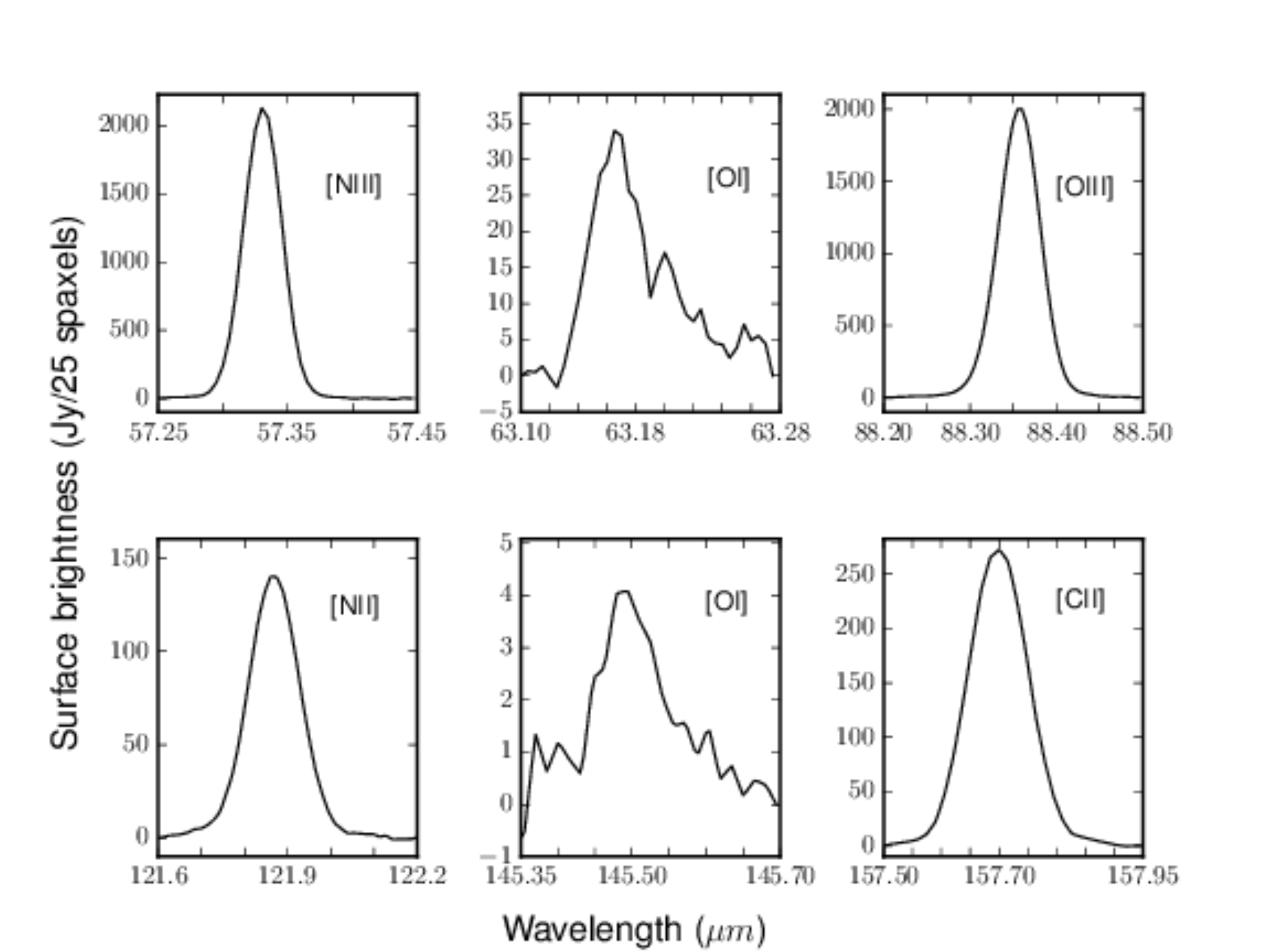}
\caption{PACS spectral lines of the H1 region in the nebula around HD\,148937, integrated over the 25 spaxels and background corrected, using the off-source region. The continuum has been removed for clarity. }\label{fig:PACS_off_dust}
\end{figure}

\begin{figure}
\centering 
\includegraphics[width=9cm,bb=5 0 537 408,clip]{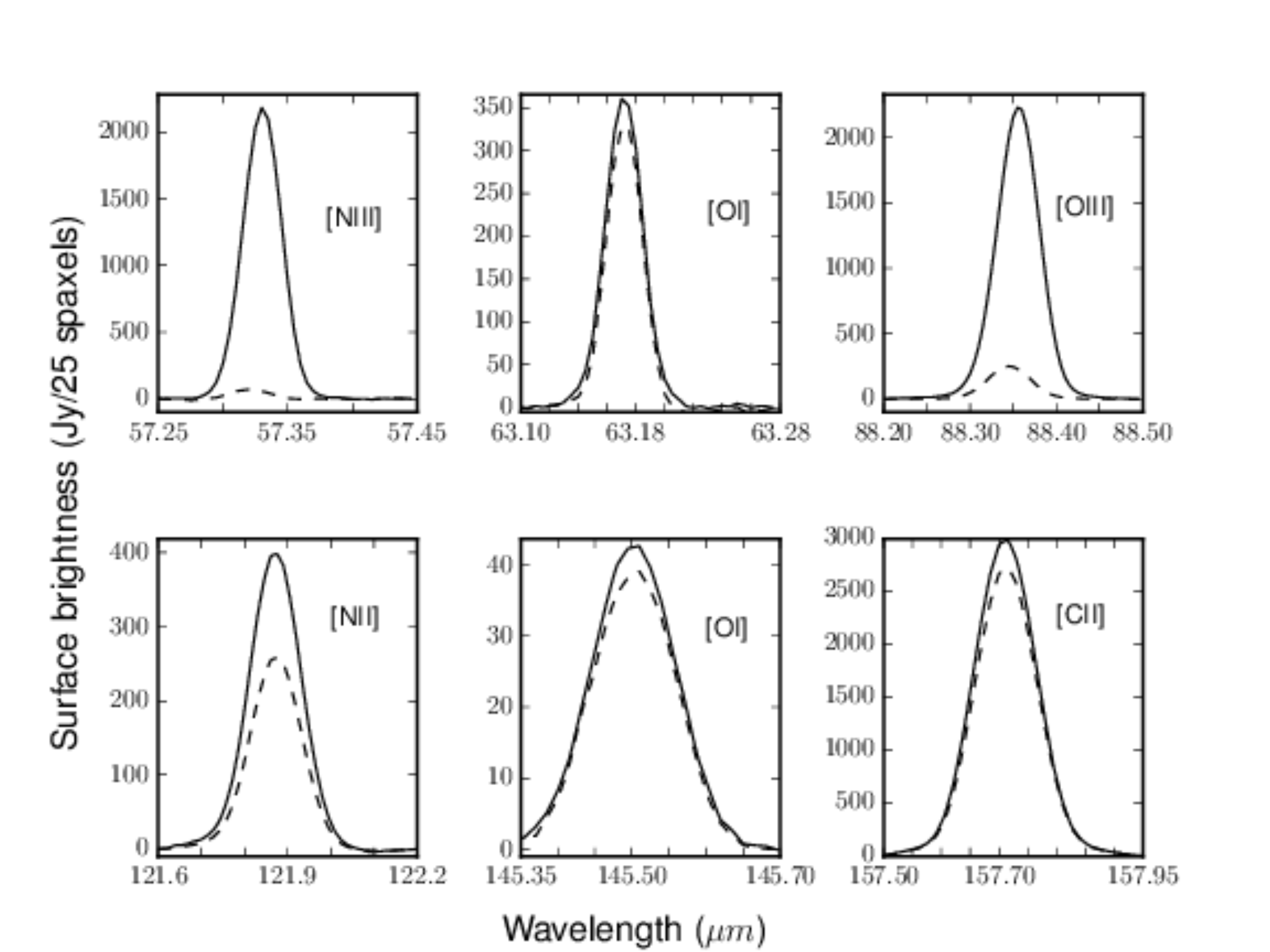}
\caption{PACS spectral lines of the H1 region (in solid line) and of the off-source region (in dashed line) integrated over the 25 spaxels. The continuum has been removed for clarity.}\label{fig:PACS_back_dust}
\end{figure}

\begin{figure}
\centering 
\includegraphics[width=9cm,bb=5 0 537 408,clip]{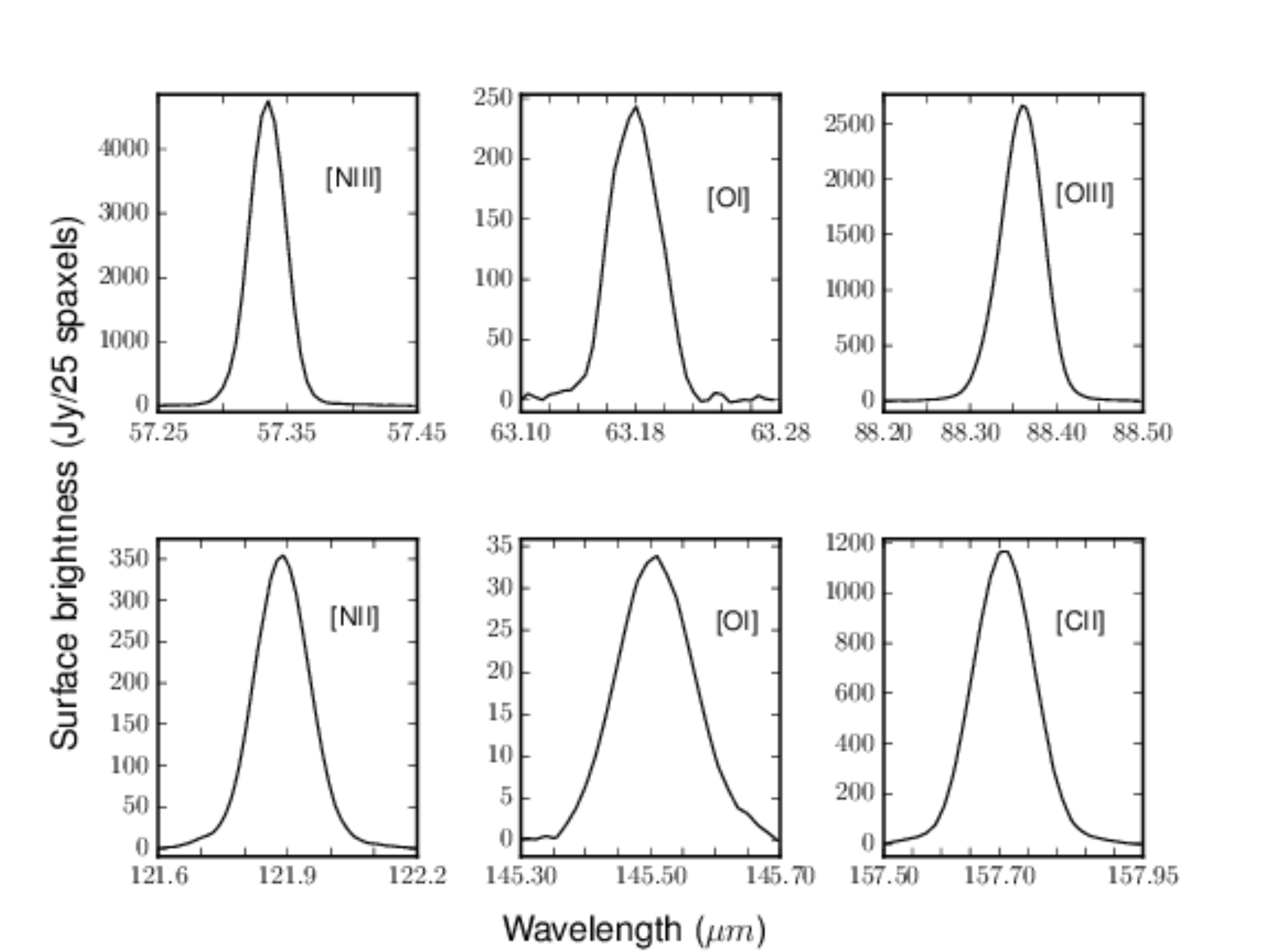}
\caption{As for Fig.\,\ref{fig:PACS_off_dust} but for the H2 region. }\label{fig:PACS_off_gas}
\end{figure}

\begin{figure}
\centering 
\includegraphics[width=9cm,bb=5 0 537 408,clip]{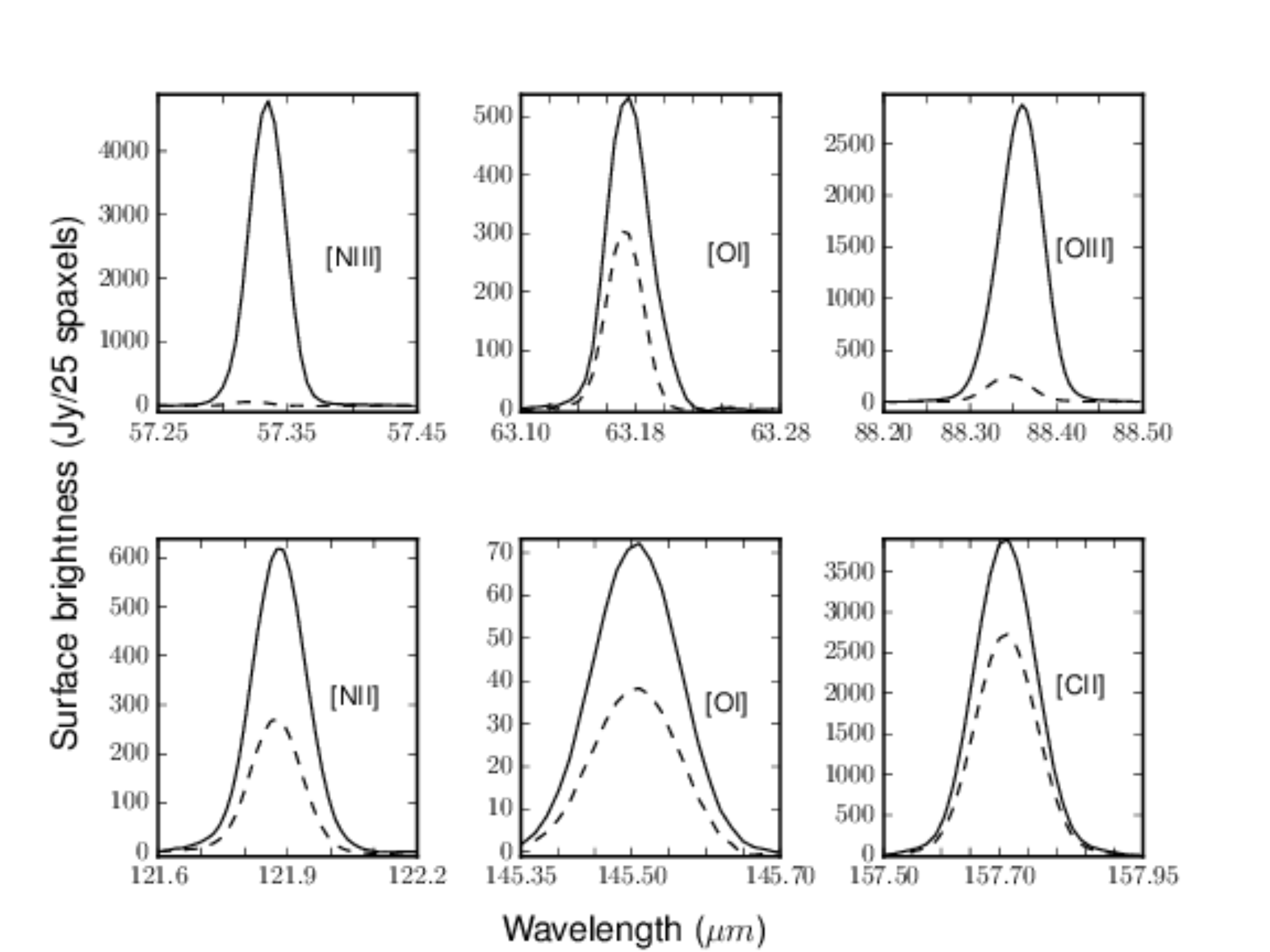}
\caption{As for Fig.\,\ref{fig:PACS_back_dust} but for the H2 region.}\label{fig:PACS_back_gas}
\end{figure}

Given the size of the nebula, only two areas of the nebula (reported as H1 and H2 in Fig.\,\ref{fig:nomen}) have been observed by PACS. The footprints of the PACS spectral field of view on the nebula as well as the off-source position are shown in Fig.\,\ref{fig:nomen}. Each PACS spectral field of view is composed of 25 ($5 \times 5$) spaxels, each corresponding to a different part of the nebula.  We measure the emission line intensities, in each one of the 25 spaxels present in the two on-source, as well as in the off-source spectra (see the tables in Appendix), by fitting a Gaussian profile to every line profile. We also measure the intensities of these lines on the integrated spectra over the 25 spaxels. The final values of the line intensities, measured on these integrated spectra correspond to the subtraction between the on-source and the off-source emission line fluxes. They are given in Table~\ref{tab:flux}. The resulting spectra are displayed in Figs.\,\ref{fig:PACS_off_dust} and\,\ref{fig:PACS_off_gas} for the H1 and for the H2 regions, respectively, whilst the comparisons between the on-source and the off-source spectra are displayed in Figs.\,\ref{fig:PACS_back_dust} and\,\ref{fig:PACS_back_gas} for the H1 and the H2 regions, respectively.

The following forbidden emission spectral lines are detected in both spectra: [\ion{N}{iii}]~57\,\mum, [\ion{O}{i}]~63\,\mum, [\ion{O}{iii}]~88\,\mum, [\ion{N}{ii}]~122\,\mum, [\ion{O}{i}]~146\,\mum, [\ion{C}{ii}]~158\,\mum\ and [\ion{N}{ii}]~205\,\mum. The presence of high ionization lines indicates that the nebula around HD\,148937 is highly ionized whilst the lowest ionization lines seem to reveal the existence of shocks or of a photodissociation region (PDR) around the central star.

\begin{table}
\caption{Line fluxes measured on the integrated spectra of the H1 and H2 regions of NGC\,6164/5, after subtracting the off-source values.}             
\label{tab:flux}      
\centering                          
\begin{tabular}{l c c}        
\hline\hline                 
Line & \multicolumn{2}{c}{Surface brightness}  \\    
& \multicolumn{2}{c}{$\times 10^{-15}$} \\
& \multicolumn{2}{c}{[W m$^{-2}/25$ spaxels]}\\
\hline                      
                 &     H1 region    &  H2 region    \\
\hline         
\protect[\ion{N}{iii}]\,57\,\mum  &   $70.9 \pm 0.2 $        &   $150.5  \pm  0.6$      \\
\protect[\ion{O}{i}]\,63\,\mum    &    $0.9 \pm  0.6$          &    $ 7.4  \pm  0.3$      \\
\protect[\ion{O}{iii}]\,88\,\mum  &   $45.8 \pm  0.2$         &     $62.3  \pm  0.1$     \\
\protect[\ion{N}{ii}]\,122\,\mum   &    $4.2 \pm  0.1$         &    $ 11.4 \pm  0.3$       \\
\protect[\ion{O}{i}]\,146\,\mum    &    $0.1 \pm 0.1 $        &    $ 0.8 \pm  0.1 $       \\
\protect[\ion{C}{ii}]\,158\,\mum   &    $4.6 \pm  0.3$        &   $ 19.3  \pm  0.4$      \\
\protect[\ion{N}{ii}]\,205\,\mum   &   $1.8  \pm 0.3 $      &      $ 3.3 \pm  0.2$       \\
\hline                                   
\end{tabular}
\end{table}

\subsubsection{The H1 region}
\paragraph{Electron density}

As we already mentioned in Sect.\,\ref{elec}, the electron temperature has been assumed to be 7000\,K through the entire nebula. To determine the electron density in the H1 region, we use the nebular/IRAF package. We compute a ratio of $2.33 \pm 0.53$ between the [\ion{N}{ii}]~122\,\mum\ and the [\ion{N}{ii}]~205\,\mum\ lines on the integrated spectrum (Table\,\ref{tab:flux}). In our spectral range, these lines are good indicators of the electron density. Assuming an electron temperature of 7000\,K, we determine an electron density of $n_e = 65 \pm 25$~cm$^{-3}$. We note that $n_e$ increases to 70~cm$^{-3}$ if we consider the upper limit on the electron temperature (16600~K) reported by \citet{leitherer87} for the H1 region. This remains, however, within the error bars. The electron density found in the H1 region is in agreement with the value obtained from the {\it Spitzer} data at position \#6. 

\paragraph{H$\alpha$ flux}
\label{Halpha_off_dust}

To determine the H$\alpha$ fluxes across the nebula, we use the H$\alpha$ image taken with Gemini, recalibrated using the {\it HST} data (see Sect\,\ref{sec:Obs}). We use the photometric aperture of the same size as the PACS spectrometer field of view. The reddened flux measured in the H1 region is estimated to $F_0($H$\alpha) = 1.06 \times 10^{-11}$~erg~cm$^{-2}$~s$^{-1}$. By assuming $(B-V) = 0.343$ \citep{maizapellaniz04} and $(B-V)_0 = -0.27$ (given the spectral type O\,5.5f?p of HD\,148937, \citealt{martins06}), we compute $E(B-V) = 0.61$. From this value, we determine de-reddened fluxes of $F_0($H$\alpha) = 4.38 \pm 0.5 \times 10^{-11}$~erg~cm$^{-2}$~s$^{-1}$ using the extinction law of \citet{cardelli89}.

\paragraph{Nebular abundances}

To compute the N$^{++}$/O$^{++}$ ratio, it requires to determine the volume emissivities of the [\ion{N}{iii}]~57\,\mum\ and the [\ion{O}{iii}]~88\,\mum\ lines. In this context, we use the nebular/IRAF package by assuming $n_e=65$~cm$^{-3}$ and $T_e = 7000$~K (see Eq.~4 in \citealt{vamvatira16}). From the measured line fluxes and their uncertainties (Table~\ref{tab:flux}), the N$^{++}$/O$^{++}$ abundance ratio is $1.06 \pm 0.04$. In order to see whether this ratio is representative of the overall N/O ratio, we determine the N$^{++}$/N$^{+}$ ratio from the [\ion{N}{iii}]~57\,\mum\ and the [\ion{N}{ii}]~122\,\mum\ lines. This ratio is equal to 2.9. As mentioned by \citet{stock14}, when this value is larger than unity, this indicates that N$^{++}$ is the dominant ionization state of nitrogen. Therefore, given the similarity of their ionization potential, we can assume that O$^{++}$ is also the dominant ionization state for oxygen in the observed field. In conclusion, we can assume that the N$^{++}$/O$^{++}$ is representative of the global N/O ratio. This value is much higher than the solar value, estimated to be $(\mathrm{N/O})_{\odot} \sim 0.14$ \citep{grevesse10}.

An estimate of the nitrogen abundance in the H1 region can also be made. In this context, we use the H$\alpha$ flux that we have determined, as well as the fluxes measured on the [\ion{N}{iii}]~57\,\mum, the [\ion{N}{ii}]~122\,\mum, and the [\ion{N}{ii}]~205\,\mum\ lines. We assume a case-B recombination with $T_e = 7000$~K \citep{draine11}. The ionic abundances N$^{+}$/H$^{+}$ and N$^{++}$/H$^{+}$ have been determined with the nebular/IRAF package, and then summed to give the final value of $\mathrm{N/H}=5.4 \pm 1.4\times 10^{-4}$, by number. If we consider $T_e = 16600$\,K (the upper limit given by \citealt{leitherer87}) and $n_e = 70$\,cm$^{-3}$, the N/H ratio is equal to $3.1 \pm 1.2\times 10^{-4}$ by number. 

\paragraph{Photodissociation region}
\label{sec:pdr}

The nebular spectrum taken in the H1 region displays the [\ion{O}{i}]~63\,\mum, the [\ion{O}{i}]~146\,\mum\ and the [\ion{C}{ii}]~158\,\mum\ lines that probably indicate the presence of a PDR around HD\,148937. In PDRs, the ratio between the [\ion{O}{i}]~63\,\mum\ and the [\ion{C}{ii}]~158\,\mum\ lines is measured to be smaller than ten \citep{tielens85}, whilst in shock regions, it is larger. Given the ratio between these two lines, it is unlikely that their presence is due to shocks.

Neutral oxygen is found only in neutral zones. Therefore, the [\ion{O}{i}] lines exclusively arise from the PDR \citep{malhotra01} but neutral atomic carbon has an ionization potential lower than that of hydrogen and one thus expects to find \ion{C}{ii} emission arising from both the \ion{H}{ii} regions and the PDR.

The C/O abundance ratio can be estimated from the PDR line fluxes following a method described by \citet{vamvatira13}. This method was used to disentangle the contributions of the PDR and of the \ion{H}{ii} region to the flux of [\ion{C}{ii}]~158\,\mum.

\begin{figure*}
\sidecaption
\includegraphics[width=12cm,bb=25 3 538 402,clip]{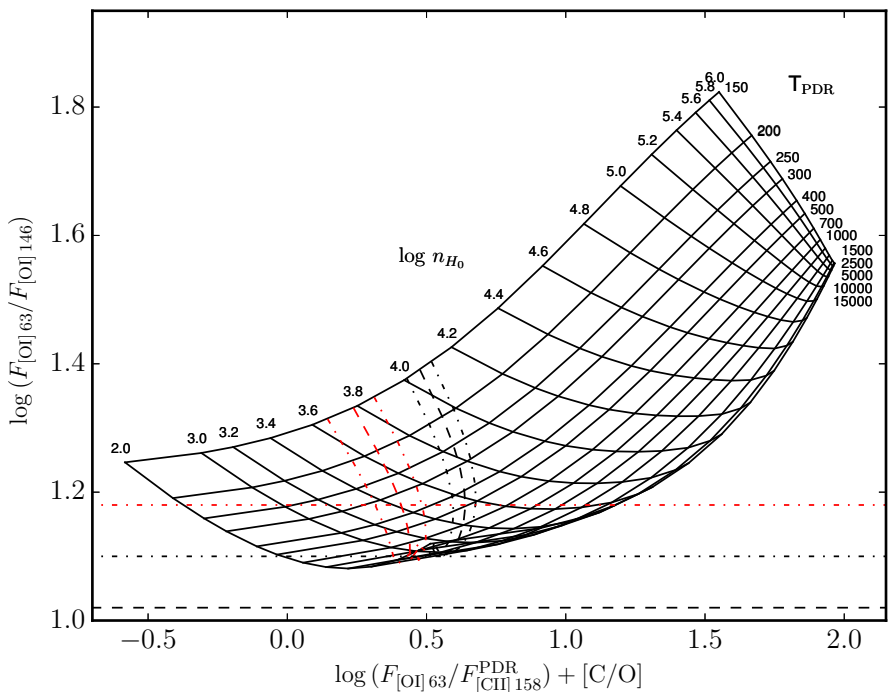}
\caption{Temperature-density PDR diagnostic diagram. The grid of flux ratios $F_{[\ion{O}{i}]~63}/F_{[\ion{O}{i}]~146}$ versus $F_{[\ion{O}{i}]~63}/F^{\mathrm{PDR}}_{[\ion{C}{ii}]~158}$ was calculated by solving the level population equations for a range of temperatures and densities. The black and the red lines correspond to the H2 and the H1 regions, respectively. The quasi-vertical dashed lines indicate the pressure equilibrium constraint between the \ion{H}{ii} region and the PDR, with the corresponding errors (dash-dot lines). The horizontal dashed line corresponds to the observed ratio with the errors \citep[see][for more details]{vamvatira13}.} \label{fig:PDRdiag}
\end{figure*}

In the ionized gas region, the ratio of fractional ionization is given by
\begin{equation}
  \frac{<\mathrm{C}^+>}{<\mathrm{N}^+>} = \frac{F^{\ion{H}{ii}}_{[\ion{C}{ii}]~158}/\epsilon_{[\ion{C}{ii}]~158}}{F_{[\ion{N}{ii}]~122}/\epsilon_{[\ion{N}{ii}]~122}}
\end{equation}
where we define $F^{\ion{H}{ii}}_{[\ion{C}{ii}]~158}=\alpha \, F_{[\ion{C}{ii}]~158}$ with $ F_{[\ion{C}{ii}]~158}$ being the total flux of the [\ion{C}{ii}]~158\,\mum\ line, $\alpha$ a factor to be determined and $\epsilon$ the volume emissivity. Assuming that $<\mathrm{C}^+>/<\mathrm{N}^+> = $ C/N and calculating the emissivities using the nebular/IRAF package with $n_e = 65$~cm$^{-3}$ and $T_e = 7000$~K, we find
\begin{equation}
  \frac{F^{\ion{H}{ii}}_{[\ion{C}{ii}]~158}}{F_{[\ion{N}{ii}]~122}} = [0.69 \pm 0.18] \frac{C}{N}.
\end{equation}
From $\mathrm{N/O}=1.06$, and from the line fluxes reported in Table\,\ref{tab:flux}, we obtain
\begin{equation}
  \log \alpha = \log~\mathrm{C/O} - [0.18 \pm 0.04],
\end{equation}
using the measured $F_{[\ion{C}{ii}]}/F_{[\ion{N}{ii}]}$ ratio.
Assuming that there is a pressure equilibrium between the ionized gas region and the PDR, we have
\begin{equation}
n_{H_0}~k~T_{\mathrm{PDR}} \cong 2~n_e~k~T_e = 9.1 \pm 0.9 \times~10^5 \mathrm{cm}^{-3} \mathrm{K}
\end{equation}
that is used to define a locus of possible values in the diagram of Fig.~\ref{fig:PDRdiag}.
From these values (red lines in Fig.~\ref{fig:PDRdiag}), we have $\log (F_{[\ion{O}{i}]~63}/F^{\mathrm{PDR}}_{[\ion{C}{ii}]~158}) + [\mathrm{C/O}] = [0.43 \pm 0.03] $, where by definition $[\mathrm{C/O}] = \log(\mathrm{C/O}) - \log(\mathrm{C/O})_{\odot}$. Using the values of the line fluxes from Table~\ref{tab:flux} and $(\mathrm{C/O})_{\odot} = 0.5$ \citep{grevesse10}, we solve the equations to obtain $\alpha = 0.95 \pm 0.03$ and $\mathrm{C/O} = 1.42 \pm 0.39 $. That means that almost all the flux of the [\ion{C}{ii}]~158\,\mum\ line comes from ionized gas. We must emphasize, however, that the estimate of the flux for the [\ion{C}{ii}]~158\,\mum\ line in the two areas observed by PACS must be taken with caution, given the high flux measured for this line in the off-source spectrum. 

\subsubsection{The H2 region}
\paragraph{Electron density}
Among the different regions of the nebula studied by \citet{leitherer87} and \citet{dufour88}, three are close to the H2 area observed by {\it Herschel}/PACS. Assuming $T_e=7000$\,K through the H2 region, as we did for the entire nebula, is in agreement with the measurements of \citet{leitherer87} and \citet{dufour88} who estimated $T_e$ to be equal to 6800K and 7600K, respectively.
We compute a ratio of $3.46 \pm 0.19 $ between the [\ion{N}{ii}]~122\,\mum\ and the [\ion{N}{ii}]~205\,\mum\ lines. To determine the electron density inside this H2 region, we use the nebular package of the IRAF/STSDAS environment. From $T_e=7000$~K and the ratio of 3.46, we determine an electron density of $n_e = 125 \pm 15$~cm$^{-3}$. This value is smaller than what we found from {\it Spitzer} data at position \#7. The aperture size and a large gradient of electron density in this region of the nebula can explain such a difference. 

\paragraph{H$\alpha$ flux}

We use the H$\alpha$ image and the same extinction value as reported in Section~\ref{Halpha_off_dust} to derive the H$\alpha$ flux in the H2 region. We determine a reddened and a de-reddened flux of $F_0($H$\alpha) = 3.32 \times 10^{-11}$~erg~cm$^{-2}$~s$^{-1}$ and $F_0($H$\alpha) = 1.46 \pm 0.6 \times 10^{-10}$~erg~cm$^{-2}$~s$^{-1}$, respectively. The values for the H$\alpha$ flux in the H2 region are therefore larger than in the H1 region. 

\paragraph{Nebular abundances}
As we did for the H1 region, we estimate the volume emissivities of the [\ion{N}{iii}]~57\,\mum\ and the [\ion{O}{iii}]~88\,\mum\ lines by assuming that $n_e=125$~cm$^{-3}$ and $T_e = 7000$~K. From these values and the fluxes reported in Table~\ref{tab:flux}, we compute a N$^{++}$/O$^{++}$ ratio of $1.54 \pm 0.06$ for the H2 region. We also compute a N$^{++}$/N$^{+}$ ratio of 1.9, indicating that the N$^{++}$/O$^{++}$ is representative of the global N/O ratio. Therefore, we conclude that the N/O value is larger than that determined for the H1 region.  Furthermore, we note that the N/O ratio estimated in the H2 region is similar to the value provided by \citet{dufour88} ($\mathrm{N/O} \sim 1.5$, for their "A'" and "B'" regions), given the uncertainties.

The nitrogen content is also computed by assuming a case-B recombination with $T_e = 7000$~K. The sum of the ionic abundances N$^{+}$/H$^{+}$ and N$^{++}$/H$^{+}$ gives a final value of $\mathrm{N/H}=4.3 \pm 1.3 \times 10^{-4}$, by number. The N/H ratio seems similar in the H1 and the H2 regions, within the uncertainties. This means that the oxygen is responsible for the difference between the N/O ratios obtained in the H1 and H2 regions, confirming its possible depletion in the nebula \citep{dufour88}.

\paragraph{Photodissociation region}

As already detected in the H1 region, the [\ion{O}{i}]~63\,\mum, [\ion{O}{i}]~146\,\mum\ and [\ion{C}{ii}]~158\,\mum\ lines are also visible in the H2 region.
By following the same method, we determine the percentage of flux arising from the ionized region for the [\ion{C}{ii}]~158\,\mum\ line as well as the C/O ratio.

In this context, we use the N/O ratio of $1.54 \pm 0.06$ that was measured in the H2 region. The temperature-density PDR diagnostic diagram (see the black lines in Fig.\,\ref{fig:PDRdiag}) provides us $\log (F_{[\ion{O}{i}]~63}/F^{\mathrm{PDR}}_{[\ion{C}{ii}]~158}) + [\mathrm{C/O}] = [0.55\pm 0.02] $, obtained from the ratio between the [\ion{O}{i}]~63\,\mum\ and the [\ion{O}{i}]~146\,\mum\ lines of about $1.0\pm 0.1$ (Table~\ref{tab:flux}). By solving the equations as reported in Sect.~\ref{sec:pdr}, we obtain $\alpha = 0.88 \pm 0.01$ and $\mathrm{C/O} = 2.24 \pm 0.10 $. The [\ion{C}{ii}] lines measured from the H1 and the H2 regions are thus clearly dominated by the \ion{H}{ii} region. Given the size of the Str{\"o}mgren radius ($\sim 1.1\degr$) surrounding the nebula NGC\,6164/5, it therefore appears normal that the PDR is only a small fraction of this immense sphere. Although the C/O ratio must be taken with caution given the background spectrum, the $\mathrm{C/O}= 1.42$ and $2.24$ in the H1 and H2 regions, respectively, can also suggest an oxygen depletion that has been observed in the nebula \citep[see, e.g.][for further details]{dufour88} and already mentioned before.

\subsubsection{Mass of the ionized nebula}

\citet{vamvatira13} provided two equations to determine the ionized mass of a nebula: one using the H$\alpha$ flux and one using the radio emission.

The H$\alpha$ flux is measured from the Gemini H$\alpha$ image, after having recalibrated it with the {\it HST} image (see Sect.\,\ref{subsubsec:gemini}). We measure a reddened flux of $F_0($H$\alpha) = 5.15 \pm 0.95 \times 10^{-10}$~erg~cm$^{-2}$~s$^{-1}$. This value is compatible, at 3 $\sigma$, with the H$\alpha$ flux estimated by \citet{frew13}. After having corrected it for the extinction, we obtain a de-reddened H$\alpha$ flux for the whole nebula of $F_0($H$\alpha) = 2.12 \pm 0.39 \times 10^{-9}$~erg~cm$^{-2}$~s$^{-1}$. The equations B.7 and B.16 provided by \citet{vamvatira13} are dependent on the fraction of the volume of the nebula that is filled by ionized gas (the $\epsilon$ parameter) as well as the fraction of He in the form of He$^{+}$. The former parameter is unknown for NGC\,6164/5 whilst the latter has been estimated to 0.14 by \citet{leitherer87}. Considering $T_e = 7000$\,K, an angular radius of the nebula equal to $185\arcsec$, and a distance of 1.3\,kpc \citep{herbst77}, we estimate the mass of the ionized gas to be $16.0 \times \sqrt{\epsilon}$\,\msun. Therefore, for an arbitrary relatively small value of the filling factor, $\epsilon = 0.01$, the ionized gas mass is 1.6\,\msun. A more conservative estimation would be to consider that all the ionized material is located in the lobe regions. In this case, we consider a shell as first approximation of the 3D ionized nebula with an outer radius of $190\arcsec$ ($\sim 1.2$ pc at 1.3~kpc) and an inner radius of $142\arcsec$ ($\sim 0.9$ pc at 1.3~kpc). Under this assumption, we estimate a volume of 4.4~pc$^3$ for the shell and of 7.4~pc$^3$ for the sphere taking the whole structure into account, giving a fraction of the volume of the ionized gas of 0.6. This value would suggest a mass of the ionized gas of 12.4\,\msun. 

The radio emission used to determine the mass of the ionized gas is given by \citet{milne82}. These authors reported an integrated flux of 2.50 Jy at $\nu=5$\,GHz and of 2.54 Jy at $\nu=14.7$\,GHz. We thus estimate the mass of the ionized gas to be $19.8 \times \sqrt{\epsilon}$\,\msun\ with the same assumptions as before. Both values are in agreement within the errors on the different parameters.

To summarize, the exact determination of the ionized mass is difficult given the uncertainties on several parameters. We can certainly assume that the mass ejected by HD\,148937 is larger  than 1.6\,\msun, which agrees with the results of \citet{leitherer87}.

   \section{Modeling of the central star}
   \label{sec:Modelling}

      To perform the modeling of the central star, we employ the non-LTE CMFGEN atmosphere code \citep{hillier98}. This code simultaneously solves the radiative transfer equation for a spherically symmetric wind in the co-moving frame and the rate equations under the constraints of radiative and statistical equilibrium. The velocity structure is constructed from a pseudo-photospheric structure connected to a $\beta$-velocity law of the form $v = \vinf (1-R/r)^{\beta}$ where \vinf\ is the terminal velocity. The density structure is computed from mass conservation. The photospheric structure is obtained after a few iterations of the hydrodynamical solution in which the radiative force computed from the level populations and atomic data is  included.

   Our final model includes \ion{H}{i}, \ion{He}{i-ii}, \ion{C}{ii-iv}, \ion{N}{ii-v}, \ion{O}{ii-vii}, \ion{Ne}{ii-iv}, \ion{Si}{iii-iv}, \ion{Mg}{ii}, \ion{S}{iii-v}, \ion{Ar}{iii-v}, \ion{Ca}{iii-iv}, \ion{Fe}{ii-vii}, and  \ion{Ni}{ii-vi} with the solar composition of \citet{grevesse10} unless otherwise stated. CMFGEN also uses the super-level approach to reduce the memory requirements. On average, we include about 1870 super levels for a total of 8050 levels. The final synthetic spectrum is obtained from a formal solution of the radiative transfer equation. A microturbulent velocity varying linearly with velocity from 10\,\kms\ to $0.1 \times \vinf$ was used. We include X-ray emission in the wind since this can affect the ionization balance and the strength of key UV diagnostic lines. In practice, we adopt a temperature of three million degrees and we adjust the flux level so that the X-ray flux coming out of the atmosphere matches the observed $L_X/L_{\mathrm{bol}}$ ratio equal to $10^{-6.1}$ \citep{naze08b}.

   The final spectrum is then convolved by a rotational profile estimated by the Fourier transform method \citep{simondiaz07}, and then convolved again by a Gaussian profile to mimick an isotropic macroturbulence. A description of the main diagnostic lines is provided by \citet{mahy15}.

   \begin{figure*}
     \centering
     \includegraphics[bb=20 0 541 406,clip]{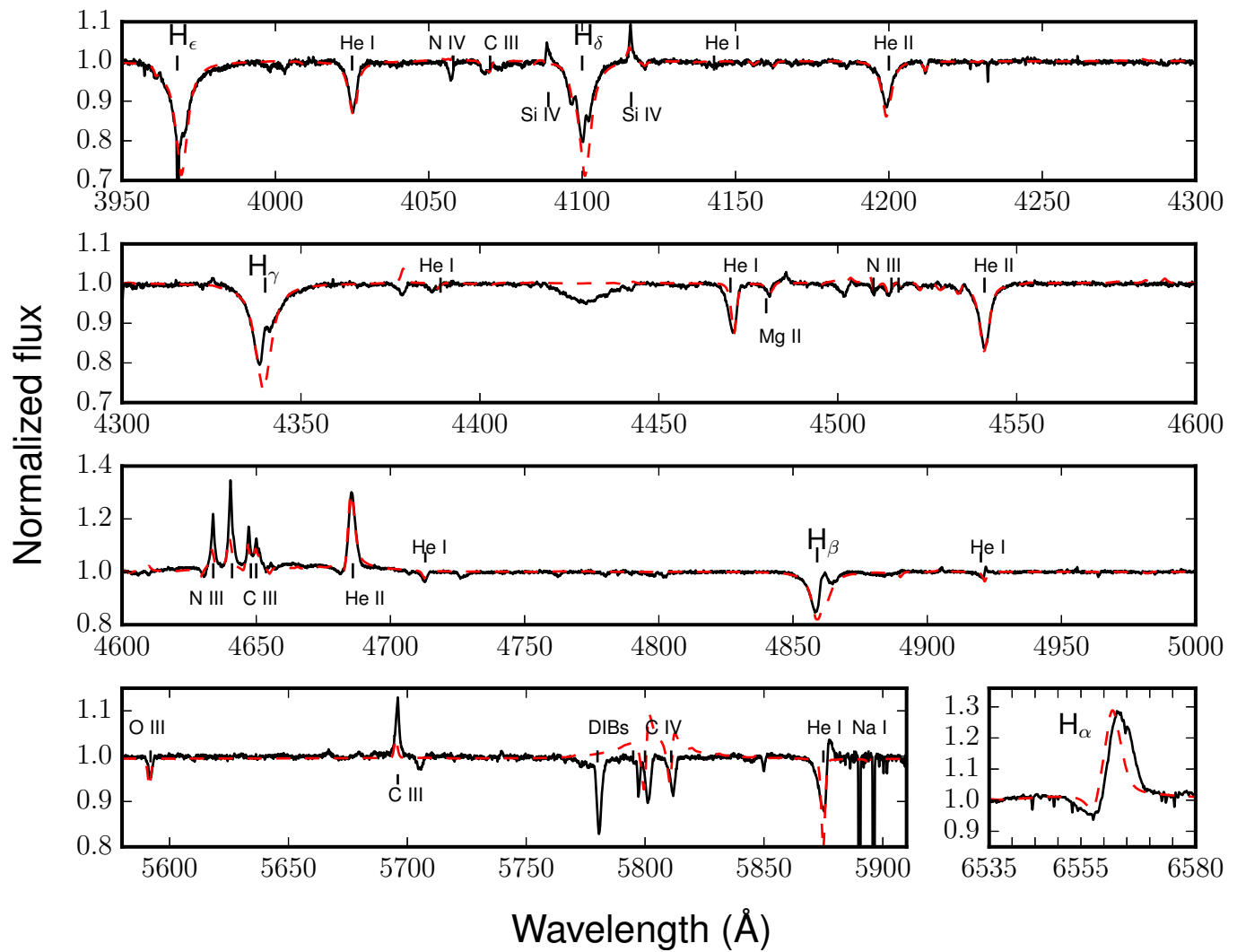}
     \caption{Best-fit CMFGEN spectrum of HD\,148937 (red line) compared to FEROS spectrum taken on June 26th, 2005. The lines that are
not modeled are diffuse interstellar bands (DIBs). Note the presence of an emission component in H and \ion{He}{i} (narrow) lines, due to confined winds. }\label{fig:CMFGEN}%
   \end{figure*}
   
   The best-fit CMFGEN spectrum (Fig.~\ref{fig:CMFGEN}) gives an effective temperature of $40000 \pm 2000$~K, and \logg\ of $4.0 \pm 0.1$. The luminosity of HD\,148937 is taken from \citet{wade12}, that is, $\lL = 5.8 \pm 0.1$. The helium lines are pretty well reproduced as well as the other absorption lines except for the cores of the Balmer lines that seem affected by emission. Only the \ion{N}{iv}~4058\AA\ and the \ion{C}{iv}~5801--11\AA\ doublet are poorly fitted. The surface carbon abundance, by number, is estimated to $2.1 \pm 0.5 \times 10^{-4}$, the surface nitrogen abundance to $4.5 \pm 0.8  \times 10^{-4}$ and the surface oxygen abundance to $1.7 \pm 0.3 \times 10^{-4}$. HD\,148937 is thus depleted in carbon and oxygen and overabundant in nitrogen. The N/O and C/O ratios determined for HD\,148937 are $2.65 \pm 1.14$ and $1.24 \pm 0.62$, respectively. N/O is higher than in the nebula, as expected qualitatively. Our stellar parameters as well as our abundances agree with the values of \citet{martins15}. The wind parameters are determined from the UV lines, the \ion{He}{ii}~4686\AA\ and H$\alpha$ lines. We are, however, not able to fit the UV and the H$\alpha$ lines with a single set of wind parameters (\mdot, \vinf, clumping filling factor, and $\beta$). A possible cause can be the spectral variability of the object, but it is most probable that, because of the magnetic confinement, the spherical symmetry assumption, implicit to CMFGEN, is no longer valid. The fundamental parameters (stellar, wind, and abundances) of HD\,148937 are listed in Table\,\ref{tab:param}.

\begin{table}
\caption{Fundamental parameters of HD\,148937}             
\label{tab:param}      
\centering                          
\begin{tabular}{l c}        
\hline\hline                 
\teff\ [K] & $40000 \pm 2000$ \\
\logg\ [cgs] & $4.0 \pm 0.1$  \\
\lL   & $5.8 \pm 0.1^{a}$ \\
$R$ [$R_{\odot}$] & $16.6_{-3.2}^{+4.0}$ \\
\mdot/$\sqrt{f}$ [\myr]& $3.0 \pm 1.0 \times 10^{-6}$ \\
\vinf\ [\kms] & $2600 \pm 350$ \\
$f$   & 0.8 \\
$v_{\mathrm{cl}}$ [\kms] & 200 \\
$\beta$ & 2.0  \\
\vsini\ [\kms] & 22 \\
\vmac\ [kms] & 70 \\
He/H$_{\mathrm{star}}$ & $0.10 \pm 0.02$ \\ 
C/H$_{\mathrm{star}}$ & $2.1 \pm 0.5 \times 10^{-4}$ \\
N/H$_{\mathrm{star}}$ & $4.5 \pm 0.8 \times 10^{-4}$ \\
O/H$_{\mathrm{star}}$ & $1.7 \pm 0.3 \times 10^{-4}$ \\
\hline                                   
\end{tabular}
\tablefoot{$^{a}$: \citet{wade12}}
\end{table}
 
   \section{Discussion}
   \label{sec:Discussion}

   \subsection{Evolutionary tracks}

   HD\,148937 is a magnetic star with a 7.03-day variability that is associated to its rotational period. \citet{wade12} determined $v_{\mathrm{eq}}/v_{\mathrm{crit}} = 0.12$ and an obliquity of the magnetic field equal to $38\degr$.

   In the present study, we confirm the effective temperature and the gravity for HD\,148937. Combined to the known luminosity of the star, these parameters provide a radius of $16.6\,R_{\odot}$, giving HD\,148937 a rotational velocity of $v_{\mathrm{eq}} \lesssim 120$\,\kms\ when accounting for the period of 7.03 days. The analysis of the optical spectrum of HD\,148937 infers a projected rotational velocity of $\vsini\ \lesssim 22$\,\kms. We thus obtain an inclination of the stellar rotation axis of $11\degr$, which is consistent, within the errors, with $i \leq 30\degr$ derived from the magnetic field analysis \citep{wade12}.

\begin{figure}
\centering 
\includegraphics[width=8cm,bb=25 5 538 402,clip]{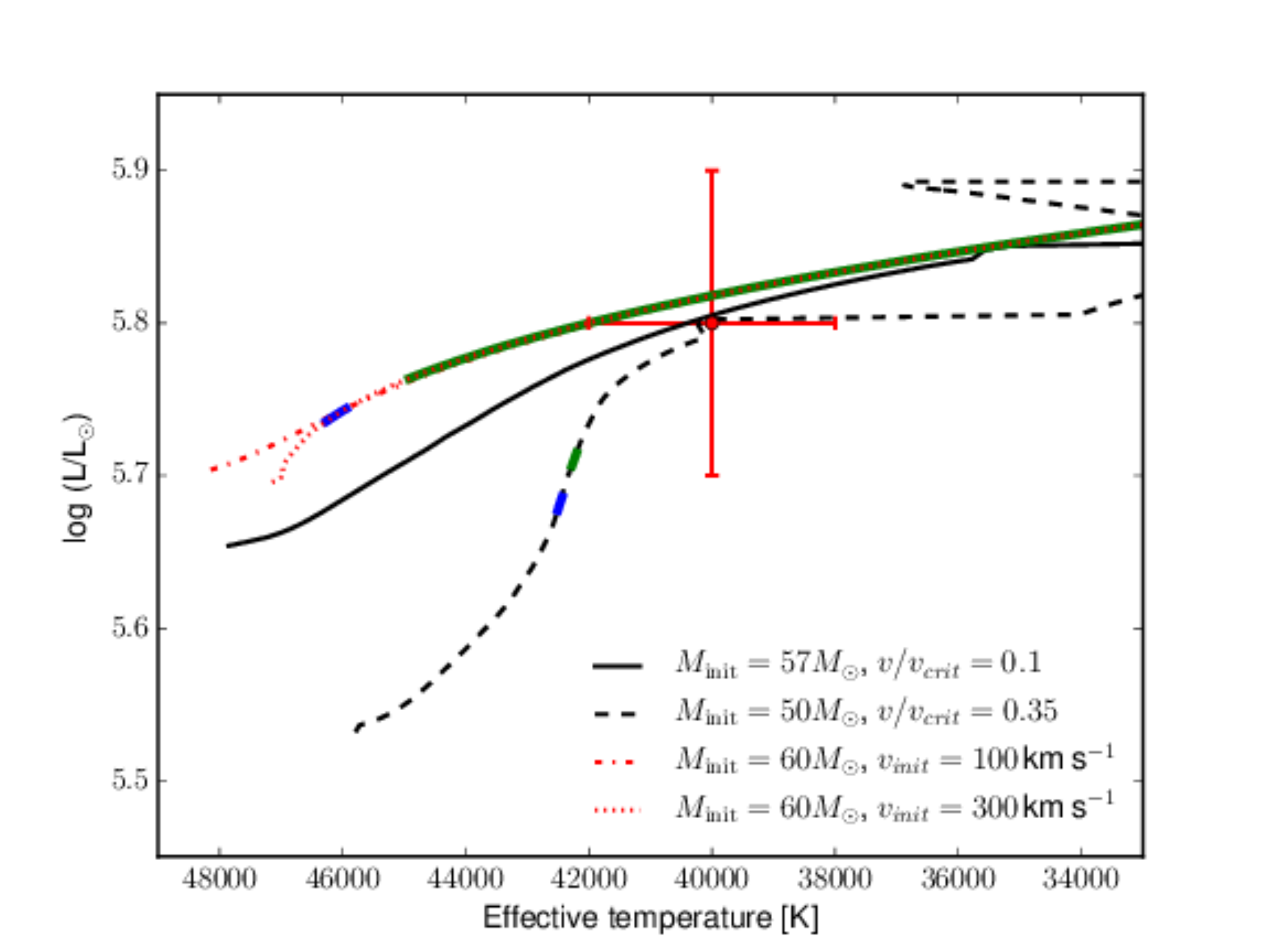}
\caption{{\it Top:} Current location of HD\,148937 (red cross) in the HR diagram and locations at the time of nebular ejections, estimated from the N/O ratios of the H1 region in blue and in the H2 region in green. Evolutionary tracks in black are from \citet{ekstrom12} whilst the red ones are from \citet{brott11}. }\label{fig:HR}
\end{figure}

   The stellar parameters of HD\,148937 allow the determination of its position in the Hertzsprung-Russell (HR) diagram (Fig.~\ref{fig:HR}). We compare this position with the evolutionary tracks of \citet{brott11} and of \citet{ekstrom12}. The evolutionary tracks of \citet{brott11} take the transport of angular momentum by magnetism into account through the Tayler-Spruit dynamo formalism \citep{spruit02,petrovic05} but do not consider possible transport of chemical elements as a result of this dynamo process, whilst the tracks of \citet{ekstrom12} do not account for magnetism, either by a dynamo mechanism or by any strong fossil field.

   The presence of a magnetic field can, however, reduce the rotational rate of the star as it was observationally detected by \citet{townsend10} in the case of $\sigma$~Ori~E. As models do not take this effect into account, we favor evolutionary tracks with low \vsini\ to constrain the current status of HD\,148937. In this context, the tracks of \citet{brott11}, computed with an initial rotational velocity of 100\,\kms, yield an initial mass of 58\,\msun\ to HD\,148937. This value agrees with the initial mass of 57\,\msun\ determined from the tracks of \citet{ekstrom12} computed with an initial rotational rate of 0.1.

   The tracks of \citet{brott11} reproduce the value of $v_{\mathrm{eq}}/v_{\mathrm{crit}} = 0.12$ at the location of HD\,148937 in the HR diagram. However, the N/O ratio that we find from the atmosphere models does not match that derived from the evolutionary tracks for a star with $\teff = 40000$\,K and $\lL=5.8$ (i.e., $\mathrm{N/O} = 0.15$). In the same way, the evolutionary tracks of \citet{ekstrom12} fail at reproducing the N/O ratio, inferring values slightly too small in comparison to the observations, as well as a value too small for the current $v_{\mathrm{eq}}/v_{\mathrm{crit}}$. It is, however, worth noting that the inclusion of magnetism, under the assumption that the Tayler-Spruit dynamo is working, in the evolutionary models predicts an increase of the surface velocity and an increase of the mixing \citep{meynet05}, implying higher surface abundances for a given initial rotational rate. An agreement can then be found between the N/O ratios derived for the nebula and the stellar value by considering tracks with an initial rotational velocity of 300\,\kms. For the tracks of \citet{ekstrom12}, the star would have an initial mass of 50\,\msun\ and would have an age of 3.8 Myrs; the material located in the H1 region with $\mathrm{N/O}=1.06$ would have been ejected by a 2.6-Myr star whilst the ejecta situated in the H2 region (where $\mathrm{N/O}=1.54$) would be produced by a 3.1-Myr star. For the tracks of \citet{brott11}, computed with an initial rotational velocity of 300\,\kms, the star would have a mass of 57\,\msun\ and currently have an age of 2.1\,Myrs, the H1 region would have been ejected when the star was 0.8\,Myrs old and the H2 region when the star was 1.6\,Myrs old. Therefore, the ejecta of the H1 region would have been ejected 1.2--1.3~Myr ago and those of the H2 region about 0.6 Myr ago. This suggests that the H2 region is younger than the H1 region even though it is further located in projection in the nebula. It thus turns out that the ejecta seem to have a complex morphology/kinematics. These values also infer an expansion velocity of about 2\,\kms which is not realistic for such an object if we consider that the material travels straight toward the lobes. All these values should however be considered as preliminary as all magnetic processes are not yet (fully) implemented in the models.  
   
   \subsection{Evolutionary scenarios}

   Up to now, HD\,148937 is the only magnetic O-type star known to be surrounded by a nebula. The exact formation process of this nebula raises many questions. The fact that HD\,148937 is, so far, the most massive Galactic star with a detected magnetic field could play a role in the presence of this circumstellar nebula even though its magnetic field is not the most powerful detected among the massive star population. Moreover, no evidence exists that HD\,148937 is a binary system. Neither \citet{naze08b} from their multiwavelength spectroscopic survey, nor \citet{sana14} from their interferometric observations have detected the presence of a companion.

   Therefore, two possible scenarios can be considered to explain the ejection of such a nebula: a giant eruption triggered by the stellar wind and the magnetic field, or a merger event between two massive stars in a binary configuration.

   \subsubsection{Giant eruption scenario}

   The ejecta constituting the nebula is known to be enriched, and has thus a stellar origin (see \citealt{leitherer87}, \citealt{dufour88} and the present study). We have roughly estimated a lower limit of 2\,\msun\ for the mass of the ejecta. \citet{garcia99} showed that the formation of planetary nebulae with such bipolar shapes is possible around magnetic stars through interactions of two succeeding time-independent stellar winds. Therefore, as a common origin for shaping bipolar nebulae, the critical rotation during a phase of strong mass loss could explain the presence of such a nebula. In this scenario, the stellar material would be first ejected in the equatorial plane. A second faster wind would then be ejected in the same plane, and would collide with the slower and denser wind and would then collimate along the polar direction. The nebula would thus be composed of two parts: an equatorial disk and a bipolar structure oriented perpendicularly.

   To see whether this scenario can be applied here, we must analyze the orientation of HD\,148937. Let us assume that the bipolar structure forming the nebula NGC\,6164/5 is oriented in the polar direction. In this case, the equatorial ejection would correspond to the inner parts detected in the different images of Fig.~\ref{fig:images}, close to the central star, whilst the lobes would be created by the collimated material. Under this assumption, the poles/lobes would be younger than the equatorial ejecta, which is what we determine from the analysis of the abundances. However, such a configuration, where the lobes are in the polar direction, cannot be reconciled with the low inclination of about $15\degr$ suggested from the 7.03-day rotational period and from the nebula (see Sect\,\ref{sec:Morpho}). It also suggests that the central star was close to its critical rotational velocity and should remain high even though the ejecta have taken out some momentum, which is also incompatible with the observations.

   Let us assume now that the bright lobes are in the equatorial plane and that the inner zones of ejecta constitute the polar structure of the nebula. This agrees better with the low inclination but, in this case, the chemical abundances determined for the lobes should be less enriched than in the inner parts of the nebula, contrary to what is observed. Furthermore, the RVs of the ejecta, close by projection to the central star, should be approaching the observer and have thus a negative value, which is also not confirmed by the different studies of the kinematics.

   \begin{figure}
\centering 
\includegraphics[width=8cm,bb=13 136 595 726,clip]{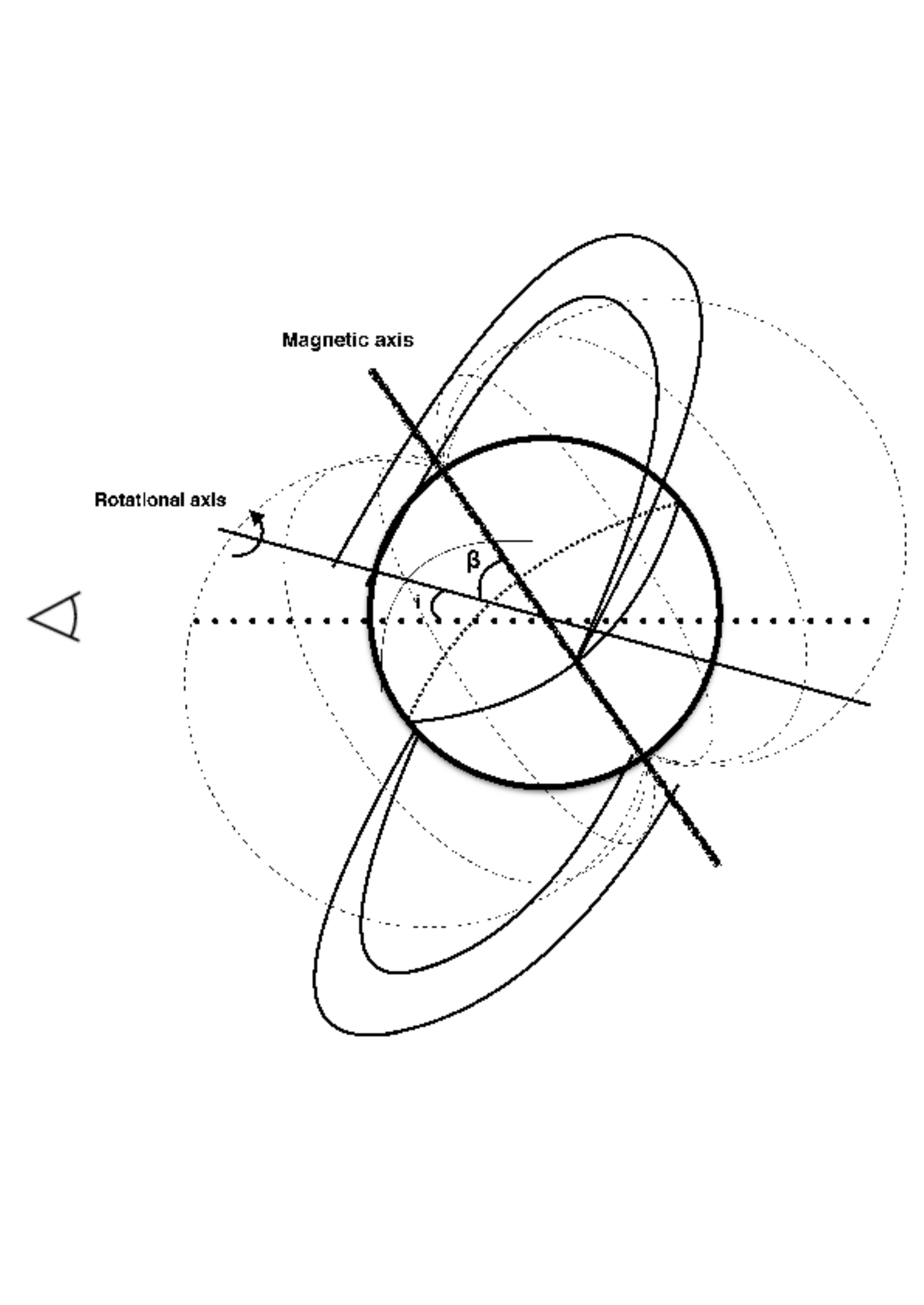}
\caption{Sketch of the helical morphology in three dimensions as viewed from the side. An Earth-based observer is at the left.}\label{fig:sketch}
\end{figure}

   The scenarios involving a giant eruption associated with wind-wind interactions thus fails at putting together all the pieces of the puzzle and at characterizing the global properties (morphology, abundances, and kinematics) of the nebula. Furthermore, by assuming that the timescales of the ejections, determined from the evolutionary models, are correct, the expansion velocity of the nebula, estimated to 2\,\kms, would be too small to justify a giant eruption. As an alternative, let us consider the morphology presented by \citet{carranza86} proposed from their kinematic analysis of the nebula. The nebula would have been ejected through a giant eruption in the equatorial plane of HD\,148937. Because of the misalignment between the magnetic and the rotational axes, the material would then follow a spiral path toward the poles of the magnetic axis, giving to the nebula a helical morphology (Fig.~\ref{fig:sketch}). The material located close to the central star by projection would have been ejected first, and thus would be less enriched than the ejecta found in the lobes, as observed. This also agrees with the low inclination of the rotational axis of HD\,148937. This scenario would turn, however, the magnetic wind channeling backwards (from the equatorial plane toward the poles) rather than from the poles toward the equational plane, as in the usual confinement scenario \citep[][and the references therein]{babel97,uddoula09}.
   
   \subsubsection{Binary merger scenario}
   
   The binary merger scenario was introduced by \citet{langer12} to explain the nature of the nebula around HD\,148937. In this interpretation, this author considered HD\,148937 as a binary merger, that may shed material in the circumstellar medium either through a bipolar nebula or through a circumbinary disk \citep[see][]{demink14}. The ages of the lobes suggested by the evolutionary tracks tend to support a less violent process, such as a merging, even though these values must still be confirmed. The merging process could even generate a magnetic field \citep{tout08} but the main issue in this scenario is the high rotational rate that would be obtained by merging. This rate would indeed be much higher than what has been derived for HD\,148937 ($v_{\mathrm{eq}} \leq 120$\,\kms). One may, however, wonder at what rate the magnetic field would brake the rotation in such a disturbed event. If this rate is very high, it could have slowed down the star quickly, but this scenario needs to be simulated to assess its validity. Another question would be to know the chemical repartition in such an event and see whether it is possible to find enriched material far from the central merger, as it is observed for HD\,148937. All these questions must still be investigated before such a scenario can be seen as physically possible and coherent considering the parameters and abundances of the star and of its ejecta obtained in the present analysis.

   \section{Conclusions}
   \label{sec:Conclusion}

   We have presented the analysis of the {\it Herschel}/PACS imaging and spectroscopic data of the nebula NGC\,6164/5 surrounding the Of?p star HD\,148937 together with H$\alpha$ and mid-infrared images, as well as high-resolution optical spectra. The H$\alpha$ image shows a bipolar or "8"-shaped ionized nebula whilst the infrared images describe a dust nebula smaller and closer to the central star. These images also exhibit a cavity in the dust material around HD\,148937 that is supposed to be created by the strong stellar wind.

   The far-infrared spectra taken in a H1 region close to the star and in a H2 region located further away, in the brightest part of NGC\,6164, were analyzed. They indicate the presence of ionized material with N/O ratios of 1.06 and 1.54, respectively. This result, combined with the studies of \citet{leitherer87} and \citet{dufour88}, shows that the ejecta forming the brightest lobes in the H$\alpha$ image are more enriched than the material located close to HD\,148937. By coupling the analysis of the abundances with the kinematic analysis of the nebula, notably presented by \citet{carranza86}, the most probable scenario capable of explaining the formation process of NGC\,6164/5 would be a giant eruption ejecting material in the equatorial plane. This material, thanks to the presence of the magnetic field, would move towards the magnetic poles through spiral motions. This scenario would suggest two different inclinations, one for the rotational axis and another one for the magnetic axis. The helical motion would thus be triggered by a cumulative effect of the stellar wind and of the magnetic field of HD\,148937. A stellar merger event can, however, not be excluded, but detailed modeling is required to see whether theoretical predictions match observations.


\begin{acknowledgements}
This research was supported by the PRODEX XMM and Herschel contract (Belspo), the Fonds National de la Recherche Scientifique (F.N.R.S.) and through the ARC grant for Concerted Research Actions, financed by the French Community of Belgium (Wallonia-Brussels Federation). We also thank Pr. G. Rauw and Dr. F. Martins for their helpful comments and Pr. D.J. Hillier for making his code CMFGEN available. We are grateful to the staffs of La Silla ESO Observatory and of Gemini Observatory for their technical support.
\end{acknowledgements}


\bibliography{hd148937_bibtex}

\newpage
\begin{appendix}

\section{Emission line fluxes for each spaxel of the H1 region}
\begin{table*}
\caption{Line fluxes in each spaxel. A dash indicates a poor signal-to-noise ratio or a non-detection.}             
\centering                          
\begin{tabular}{l |c |c |c |c |c}        
\hline\hline                 
Line & Flux  & Flux & Flux &  Flux &  Flux  \\
     & [$10^{-15}$ W m$^{-2}$] &[$10^{-15}$ W m$^{-2}$]&[$10^{-15}$ W m$^{-2}$]&[$10^{-15}$ W m$^{-2}$]&[$10^{-15}$ W m$^{-2}$]\\
\hline
&  spaxel (0,0) & spaxel (0,1) & spaxel (0,2) & spaxel (0,3) & spaxel (0,4)\\
\hline
$[\ion{N}{iii}]\,57\,\mum$ & $3.50 \pm 0.01$ & $3.60 \pm 0.01$   & $3.53 \pm 0.01 $   & $3.56 \pm 0.01 $   & $ 3.17 \pm 0.01 $ \\
$[\ion{O}{i}]\,63\,\mum$   & $0.36 \pm 0.02$ & $0.36 \pm 0.02$   & $0.35 \pm 0.02$    & $0.35 \pm 0.02 $   & $ 0.34 \pm 0.02 $ \\
$[\ion{O}{iii}]\,88\,\mum$ & $2.40 \pm 0.01$ & $2.40 \pm 0.01$   & $2.41 \pm 0.01$    & $2.68 \pm 0.01$    & $ 2.45 \pm 0.01 $ \\
$[\ion{N}{ii}]\,122\,\mum$ & $0.56 \pm 0.00$ & $0.55 \pm 0.00$   & $0.48 \pm 0.00$    & $0.45 \pm 0.00$    & $ 0.44 \pm 0.00$ \\
$[\ion{O}{i}]\,146\,\mum$  & $0.02 \pm 0.00$ & $0.03 \pm 0.00$   & $0.05 \pm 0.00$    & $0.03 \pm 0.00$    & $ 0.02 \pm 0.00$  \\
$[\ion{C}{ii}]\,158\,\mum$ & $1.91 \pm 0.01$& $2.09 \pm 0.01$  & $2.02 \pm 0.01$   & $2.02 \pm 0.01$   & $ 1.98 \pm 0.01$ \\
$[\ion{N}{ii}]\,205\,\mum$ & $0.15 \pm 0.01$ & $0.31 \pm 0.01$   & $0.22 \pm 0.01$   & $0.29 \pm 0.01$   & $ 0.20 \pm 0.01$ \\
\hline
&  spaxel (1,0) & spaxel (1,1) & spaxel (1,2) & spaxel (1,3) & spaxel (1,4)\\
\hline
$[\ion{N}{iii}]\,57\,\mum$ & $3.31 \pm 0.01$  & $3.33 \pm 0.01$   & $4.21 \pm 0.01 $   & $4.09 \pm 0.01 $   & $ 3.09 \pm 0.01 $ \\
$[\ion{O}{i}]\,63\,\mum$   & $0.29 \pm 0.02$ & $0.31 \pm 0.02$  & $0.33 \pm 0.02$   & $0.38 \pm 0.02$   & $ 0.32 \pm 0.02$ \\
$[\ion{O}{iii}]\,88\,\mum$ & $2.28 \pm 0.01$  & $2.13 \pm 0.01$   & $2.81 \pm 0.01$   & $2.76 \pm 0.01 $   & $ 2.30 \pm 0.01 $ \\
$[\ion{N}{ii}]\,122\,\mum$ & $0.57 \pm 0.00$ & $0.53 \pm 0.01$  & $0.56 \pm 0.01$   & $0.46 \pm 0.01$   & $ 0.42 \pm 0.00$ \\
$[\ion{O}{i}]\,146\,\mum$  & $0.03 \pm 0.00$ & $0.03 \pm 0.01$  & $0.03 \pm 0.01$   & $0.04 \pm 0.01$   & $ 0.02 \pm 0.00$  \\
$[\ion{C}{ii}]\,158\,\mum$ & $1.99 \pm 0.01$ & $2.04 \pm 0.01$  & $1.98 \pm 0.01$   & $1.95 \pm 0.01$   & $ 1.82 \pm 0.01$ \\
$[\ion{N}{ii}]\,205\,\mum$ & $0.21 \pm 0.01$ & $0.22 \pm 0.01$  & $0.28 \pm 0.01$   & $0.26 \pm 0.01$   & $ 0.28 \pm 0.01$  \\
\hline
&  spaxel (2,0) & spaxel (2,1) & spaxel (2,2) & spaxel (2,3) & spaxel (2,4)\\
\hline
$[\ion{N}{iii}]\,57\,\mum$ & $2.78 \pm 0.01$  & $2.76 \pm 0.01$   & $3.64 \pm 0.01 $   & $3.52 \pm 0.01 $   & $ 3.00 \pm 0.01 $ \\
$[\ion{O}{i}]\,63\,\mum$   & $0.35 \pm 0.03$  & $0.31 \pm 0.04$  & $0.31 \pm 0.05$   & $0.36 \pm 0.04$   & $ 0.35 \pm 0.03$ \\
$[\ion{O}{iii}]\,88\,\mum$ & $1.97 \pm 0.01$  & $1.85 \pm 0.01$   & $2.45 \pm 0.01 $   & $2.38 \pm 0.01 $   & $ 2.13 \pm 0.01 $ \\
$[\ion{N}{ii}]\,122\,\mum$ & $0.44 \pm 0.00$  & $0.49 \pm 0.01$  & $0.49 \pm 0.01$   & $0.53 \pm 0.01$   & $ 0.47 \pm 0.00$ \\
$[\ion{O}{i}]\,146\,\mum$  & $0.03 \pm 0.00$  & $0.05 \pm 0.01$  & $0.06 \pm 0.01$   & $0.02 \pm 0.01$   & $ 0.04 \pm 0.00$  \\
$[\ion{C}{ii}]\,158\,\mum$ & $1.88 \pm 0.01$  & $1.93 \pm 0.01$  & $1.86 \pm 0.02$   & $1.95 \pm 0.01$   & $ 1.91 \pm 0.01$ \\
$[\ion{N}{ii}]\,205\,\mum$ & $0.23 \pm 0.01$  & $0.20 \pm 0.01$  & $0.21 \pm 0.01$   & $0.27 \pm 0.02$   & $ 0.26 \pm 0.02$  \\
\hline
&  spaxel (3,0) & spaxel (3,1) & spaxel (3,2) & spaxel (3,3) & spaxel (3,4)\\
\hline
$[\ion{N}{iii}]\,57\,\mum$ & $2.45 \pm 0.01$  & $2.12 \pm 0.01$   & $2.49 \pm 0.01 $   & $2.36 \pm 0.01 $   & $ 2.74 \pm 0.01 $ \\
$[\ion{O}{i}]\,63\,\mum$   & $0.33 \pm 0.02$  & $0.28 \pm 0.02$   & $0.28 \pm 0.02$    & $0.30 \pm 0.02$    & $ 0.37 \pm 0.02$ \\
$[\ion{O}{iii}]\,88\,\mum$ & $1.84 \pm 0.01$  & $1.62 \pm 0.01$   & $1.85 \pm 0.01 $   & $1.72 \pm 0.01 $   & $ 1.83 \pm 0.01 $ \\
$[\ion{N}{ii}]\,122\,\mum$ & $0.40 \pm 0.00$ & $0.48 \pm 0.01$  & $0.44 \pm 0.01$   & $0.45 \pm 0.01$   & $ 0.46 \pm 0.00$ \\
$[\ion{O}{i}]\,146\,\mum$  & $0.03 \pm 0.00$ & $0.02 \pm 0.01$  & $0.03 \pm 0.01$   & $0.03 \pm 0.00$   & $ 0.04 \pm 0.00$  \\
$[\ion{C}{ii}]\,158\,\mum$ & $1.89 \pm 0.01$ & $1.81 \pm 0.01$  & $1.90 \pm 0.01$   & $1.93 \pm 0.01$   & $ 1.93 \pm 0.01$ \\
$[\ion{N}{ii}]\,205\,\mum$ & $0.21 \pm 0.01$ & $0.31 \pm 0.01$  & $0.31 \pm 0.01$   & $0.32 \pm 0.01$   & $ 0.15 \pm 0.02$  \\
\hline
&  spaxel (4,0) & spaxel (4,1) & spaxel (4,2) & spaxel (4,3) & spaxel (4,4)\\
\hline
$[\ion{N}{iii}]\,57\,\mum$ & $2.35 \pm 0.01$  & $1.99 \pm 0.01$   & $1.88 \pm 0.01 $   & $1.83 \pm 0.01 $   & $ 1.79 \pm 0.01 $ \\
$[\ion{O}{i}]\,63\,\mum$   & $0.37 \pm 0.02$  & $0.31 \pm 0.02$   & $0.31 \pm 0.02$    & $0.44 \pm 0.02$    & $ 0.37 \pm 0.02$ \\
$[\ion{O}{iii}]\,88\,\mum$ & $1.68 \pm 0.01$  & $1.42 \pm 0.01$   & $1.42 \pm 0.01 $   & $1.27 \pm 0.01 $   & $ 1.32 \pm 0.01 $ \\
$[\ion{N}{ii}]\,122\,\mum$ & $0.37 \pm 0.00$ & $0.43 \pm 0.00$  & $0.43 \pm 0.00$   & $0.36 \pm 0.00$   & $ 0.38 \pm 0.00$ \\
$[\ion{O}{i}]\,146\,\mum$  & $0.02 \pm 0.01$ & $0.04 \pm 0.01$  & $0.04 \pm 0.00$   & $0.06 \pm 0.01$   & $ 0.03 \pm 0.00$  \\
$[\ion{C}{ii}]\,158\,\mum$ & $1.65 \pm 0.01$ & $1.72 \pm 0.01$  & $1.91 \pm 0.01$   & $1.88 \pm 0.01$   & $ 1.90 \pm 0.01$ \\
$[\ion{N}{ii}]\,205\,\mum$ & $0.14 \pm 0.01$ & $0.12 \pm 0.01$  & $0.24 \pm 0.01$   & $0.21 \pm 0.01$   & $ 0.22 \pm 0.01$  \\
\hline                                   
\end{tabular}
\end{table*}

\section{Emission line fluxes for each spaxel of the H2 region}
\begin{table*}
\caption{Line fluxes in each spaxel. A dash indicates a poor signal-to-noise ratio or a non-detection.}             
\centering                          
\begin{tabular}{l |c |c |c |c |c}        
\hline\hline                 
Line & Flux  & Flux & Flux &  Flux &  Flux  \\
     & [$10^{-15}$ W m$^{-2}$] &[$10^{-15}$ W m$^{-2}$]&[$10^{-15}$ W m$^{-2}$]&[$10^{-15}$ W m$^{-2}$]&[$10^{-15}$ W m$^{-2}$]\\
\hline
&  spaxel (0,0) & spaxel (0,1) & spaxel (0,2) & spaxel (0,3) & spaxel (0,4)\\
\hline
$[\ion{N}{iii}]\,57\,\mum$ & $5.55 \pm 0.02$ & $8.35 \pm 0.02$   & $8.04 \pm 0.02 $   & $4.96 \pm 0.02 $   & $ 1.72 \pm 0.02 $ \\
$[\ion{O}{i}]\,63\,\mum$   & $0.53 \pm 0.01$ & $0.77 \pm 0.01$   & $0.96 \pm 0.01$    & $0.60 \pm 0.01 $   & $ 0.58 \pm 0.01 $ \\
$[\ion{O}{iii}]\,88\,\mum$ & $2.76 \pm 0.01$ & $3.24 \pm 0.01$   & $3.09 \pm 0.01$    & $2.41 \pm 0.01$    & $ 1.35 \pm 0.01 $ \\
$[\ion{N}{ii}]\,122\,\mum$ & $0.73 \pm 0.01$ & $0.90 \pm 0.01$   & $1.07 \pm 0.01$   & $0.91 \pm 0.01$   & $ 0.80 \pm 0.01$ \\
$[\ion{O}{i}]\,146\,\mum$  & $0.06 \pm 0.00$ & $0.10 \pm 0.01$   & $0.10 \pm 0.01$    & $0.06 \pm 0.01$    & $ 0.06 \pm 0.00$  \\
$[\ion{C}{ii}]\,158\,\mum$ & $2.58 \pm 0.02$& $2.82 \pm 0.02$  & $2.91 \pm 0.01$   & $2.87 \pm 0.01$   & $ 2.94 \pm 0.02$ \\
$[\ion{N}{ii}]\,205\,\mum$ & $0.19 \pm 0.01$ & $0.34 \pm 0.01$   & $0.21 \pm 0.01$   & $0.44 \pm 0.01$   & $ 0.30 \pm 0.01$ \\
\hline
&  spaxel (1,0) & spaxel (1,1) & spaxel (1,2) & spaxel (1,3) & spaxel (1,4)\\
\hline
$[\ion{N}{iii}]\,57\,\mum$ & $5.67 \pm 0.02$  & $9.34 \pm 0.02$   & $9.12 \pm 0.02 $   & $6.50 \pm 0.02 $   & $ 2.19 \pm 0.02 $ \\
$[\ion{O}{i}]\,63\,\mum$   & $0.42 \pm 0.02$ & $0.61 \pm 0.02$  & $1.52 \pm 0.02$   & $0.55 \pm 0.02$   & $ 0.53 \pm 0.02$ \\
$[\ion{O}{iii}]\,88\,\mum$ & $2.87 \pm 0.02$  & $3.52 \pm 0.02$   & $3.37 \pm 0.02 $   & $2.75 \pm 0.02 $   & $ 1.59 \pm 0.02 $ \\
$[\ion{N}{ii}]\,122\,\mum$ & $0.80 \pm 0.01$ & $0.96 \pm 0.01$  & $1.47 \pm 0.01$   & $0.92 \pm 0.01$   & $ 0.69 \pm 0.01$ \\
$[\ion{O}{i}]\,146\,\mum$  & $0.06 \pm 0.01$ & $0.06 \pm 0.01$  & $0.10 \pm 0.00$   & $0.08 \pm 0.00$   & $ 0.05 \pm 0.01$  \\
$[\ion{C}{ii}]\,158\,\mum$ & $2.62 \pm 0.01$ & $2.72 \pm 0.01$  & $2.89 \pm 0.02$   & $2.78 \pm 0.02$   & $ 2.57 \pm 0.01$ \\
$[\ion{N}{ii}]\,205\,\mum$ & $0.31 \pm 0.01$ & $0.24 \pm 0.01$  & $0.37 \pm 0.00$   & $0.36 \pm 0.00$   & $ 0.42 \pm 0.01$  \\
\hline
&  spaxel (2,0) & spaxel (2,1) & spaxel (2,2) & spaxel (2,3) & spaxel (2,4)\\
\hline
$[\ion{N}{iii}]\,57\,\mum$ & $5.46 \pm 0.04$  & $7.63 \pm 0.04$   & $10.66\pm 0.04 $   & $6.68 \pm 0.04 $   & $ 3.25 \pm 0.04 $ \\
$[\ion{O}{i}]\,63\,\mum$   & $0.45 \pm 0.01$ & $0.45 \pm 0.01$  & $0.85 \pm 0.01$   & $0.58 \pm 0.01$   & $ 0.42 \pm 0.01$ \\
$[\ion{O}{iii}]\,88\,\mum$ & $2.79 \pm 0.02$  & $3.29 \pm 0.02$   & $3.75 \pm 0.02 $   & $2.76 \pm 0.02 $   & $ 1.84 \pm 0.02 $ \\
$[\ion{N}{ii}]\,122\,\mum$ & $0.71 \pm 0.01$ & $0.70 \pm 0.01$  & $1.00 \pm 0.01$   & $0.86 \pm 0.01$   & $ 0.61 \pm 0.01$ \\
$[\ion{O}{i}]\,146\,\mum$  & $0.04 \pm 0.01$ & $0.07 \pm 0.01$  & $0.08 \pm 0.01$   & $0.04 \pm 0.01$   & $ 0.06 \pm 0.01$  \\
$[\ion{C}{ii}]\,158\,\mum$ & $2.45 \pm 0.03$ & $2.43 \pm 0.02$  & $2.49 \pm 0.02$   & $2.43 \pm 0.02$   & $ 2.51 \pm 0.01$ \\
$[\ion{N}{ii}]\,205\,\mum$ & $0.23 \pm 0.01$ & $0.27 \pm 0.01$  & $0.28 \pm 0.01$   & $0.41 \pm 0.01$   & $ 0.34 \pm 0.01$  \\
\hline
&  spaxel (3,0) & spaxel (3,1) & spaxel (3,2) & spaxel (3,3) & spaxel (3,4)\\
\hline
$[\ion{N}{iii}]\,57\,\mum$ & $4.53 \pm 0.02$  & $5.75 \pm 0.02$   & $8.55 \pm 0.02 $   & $7.99 \pm 0.02 $   & $ 4.49 \pm 0.02 $ \\
$[\ion{O}{i}]\,63\,\mum$   & $0.40 \pm 0.01$ & $0.46 \pm 0.01$  & $0.66 \pm 0.01$   & $0.61 \pm 0.01$   & $ 0.56 \pm 0.01$ \\
$[\ion{O}{iii}]\,88\,\mum$ & $2.53 \pm 0.02$  & $2.77 \pm 0.02$   & $3.44 \pm 0.02 $   & $3.02 \pm 0.02 $   & $ 2.20 \pm 0.02 $ \\
$[\ion{N}{ii}]\,122\,\mum$ & $0.57 \pm 0.01$ & $0.57 \pm 0.01$  & $0.70 \pm 0.01$   & $0.86 \pm 0.01$   & $ 0.70 \pm 0.01$ \\
$[\ion{O}{i}]\,146\,\mum$  & $0.04 \pm 0.00$ & $0.04 \pm 0.00$  & $0.07 \pm 0.01$   & $0.04 \pm 0.01$   & $ 0.05 \pm 0.00$  \\
$[\ion{C}{ii}]\,158\,\mum$ & $2.33 \pm 0.01$ & $2.34 \pm 0.01$  & $2.46 \pm 0.01$   & $2.31 \pm 0.02$   & $ 2.37 \pm 0.02$ \\
$[\ion{N}{ii}]\,205\,\mum$ & $0.30 \pm 0.01$ & $0.30 \pm 0.01$  & $0.30 \pm 0.01$   & $0.30 \pm 0.01$   & $ 0.27 \pm 0.01$  \\
\hline
&  spaxel (4,0) & spaxel (4,1) & spaxel (4,2) & spaxel (4,3) & spaxel (4,4)\\
\hline
$[\ion{N}{iii}]\,57\,\mum$ & $4.02 \pm 0.02$  & $4.94 \pm 0.02$   & $5.90 \pm 0.02 $   & $6.48 \pm 0.02 $   & $ 4.15 \pm 0.02 $ \\
$[\ion{O}{i}]\,63\,\mum$   & $0.40 \pm 0.01$ & $0.38 \pm 0.01$  & $0.37 \pm 0.01$   & $0.46 \pm 0.01$   & $ 0.42 \pm 0.01$ \\
$[\ion{O}{iii}]\,88\,\mum$ & $2.32 \pm 0.01$  & $2.47 \pm 0.01$   & $2.85 \pm 0.01 $   & $2.82 \pm 0.01 $   & $ 2.18 \pm 0.01 $ \\
$[\ion{N}{ii}]\,122\,\mum$ & $0.42 \pm 0.01$ & $0.51 \pm 0.01$  & $0.54 \pm 0.01$   & $0.55 \pm 0.01$   & $ 0.57 \pm 0.01$ \\
$[\ion{O}{i}]\,146\,\mum$  & $0.03 \pm 0.00$ & $0.04 \pm 0.00$  & $0.05 \pm 0.00$   & $0.06 \pm 0.00$   & $ 0.03 \pm 0.00$  \\
$[\ion{C}{ii}]\,158\,\mum$ & $2.08 \pm 0.01$ & $2.05 \pm 0.02$  & $2.30 \pm 0.01$   & $2.22 \pm 0.02$   & $ 2.23 \pm 0.02$ \\
$[\ion{N}{ii}]\,205\,\mum$ & $0.12 \pm 0.01$ & $0.16 \pm 0.01$  & $0.18 \pm 0.01$   & $0.33 \pm 0.01$   & $ 0.29 \pm 0.00$  \\
\hline                                   
\end{tabular}
\end{table*}

\section{Emission line fluxes for each spaxel of the off-source region}
\begin{table*}
\caption{Line fluxes in each spaxel. A dash indicates a poor signal-to-noise ratio or a non-detection.}             
\centering                          
\begin{tabular}{l |c |c |c |c |c}        
\hline\hline                 
Line & Flux  & Flux & Flux &  Flux &  Flux  \\
     & [$10^{-15}$ W m$^{-2}$] &[$10^{-15}$ W m$^{-2}$]&[$10^{-15}$ W m$^{-2}$]&[$10^{-15}$ W m$^{-2}$]&[$10^{-15}$ W m$^{-2}$]\\
\hline
&  spaxel (0,0) & spaxel (0,1) & spaxel (0,2) & spaxel (0,3) & spaxel (0,4)\\
\hline
$[\ion{N}{iii}]\,57\,\mum$ & $0.08 \pm 0.00$ & $0.13 \pm 0.01$   & $0.09 \pm 0.02 $   & $0.06 \pm 0.00 $   & $ 0.06 \pm 0.01 $ \\
$[\ion{O}{i}]\,63\,\mum$   & $0.36 \pm 0.01$ & $0.31 \pm 0.01$   & $0.33 \pm 0.01$    & $0.34 \pm 0.01 $   & $ 0.30 \pm 0.01 $ \\
$[\ion{O}{iii}]\,88\,\mum$ & $0.24 \pm 0.01$ & $0.24 \pm 0.01$   & $0.23 \pm 0.01$    & $0.30 \pm 0.01$    & $ 0.23 \pm 0.01 $ \\
$[\ion{N}{ii}]\,122\,\mum$ & $0.33 \pm 0.01$ & $0.37 \pm 0.01$   & $0.32 \pm 0.01$    & $0.32 \pm 0.01$    & $ 0.39 \pm 0.01$ \\
$[\ion{O}{i}]\,146\,\mum$  & $0.03 \pm 0.00$ & $0.03 \pm 0.00$   & $0.06 \pm 0.01$    & $0.03 \pm 0.00$    & $ 0.02 \pm 0.01$  \\
$[\ion{C}{ii}]\,158\,\mum$ & $1.76 \pm 0.01$ & $1.86 \pm 0.02$   & $1.87 \pm 0.02$    & $1.91 \pm 0.01$    & $ 1.90 \pm 0.02$ \\
$[\ion{N}{ii}]\,205\,\mum$ & $0.11 \pm 0.01$ & $0.22 \pm 0.01$   & $0.08 \pm 0.01$    & $0.15 \pm 0.01$    & $ 0.12 \pm 0.01$ \\
\hline
&  spaxel (1,0) & spaxel (1,1) & spaxel (1,2) & spaxel (1,3) & spaxel (1,4)\\
\hline
$[\ion{N}{iii}]\,57\,\mum$ & $0.05 \pm 0.01$  & $0.13 \pm 0.01$   & $0.04 \pm 0.01$   & $0.09 \pm 0.00$   & $ 0.09 \pm 0.01$ \\
$[\ion{O}{i}]\,63\,\mum$   & $0.34 \pm 0.01$  & $0.25 \pm 0.01$   & $0.31 \pm 0.02$   & $0.36 \pm 0.01$   & $ 0.32 \pm 0.01$ \\
$[\ion{O}{iii}]\,88\,\mum$ & $0.21 \pm 0.01$  & $0.27 \pm 0.01$   & $0.25 \pm 0.01$   & $0.22 \pm 0.01$   & $ 0.24 \pm 0.01$ \\
$[\ion{N}{ii}]\,122\,\mum$ & $0.33 \pm 0.01$  & $0.36 \pm 0.01$   & $0.32 \pm 0.01$   & $0.28 \pm 0.01$   & $ 0.29 \pm 0.01$ \\
$[\ion{O}{i}]\,146\,\mum$  & $0.03 \pm 0.01$  & $0.02 \pm 0.00$   & $0.02 \pm 0.00$   & $0.04 \pm 0.01$   & $ 0.02 \pm 0.01$  \\
$[\ion{C}{ii}]\,158\,\mum$ & $1.82 \pm 0.02$  & $1.66 \pm 0.02$   & $1.73 \pm 0.02$   & $1.81 \pm 0.02$   & $ 1.74 \pm 0.01$ \\
$[\ion{N}{ii}]\,205\,\mum$ & $0.12 \pm 0.01$  & $0.14 \pm 0.01$   & $0.20 \pm 0.01$   & $0.19 \pm 0.01$   & $ 0.22 \pm 0.01$  \\
\hline
&  spaxel (2,0) & spaxel (2,1) & spaxel (2,2) & spaxel (2,3) & spaxel (2,4)\\
\hline
$[\ion{N}{iii}]\,57\,\mum$ & $0.14 \pm 0.01$  & $0.11 \pm 0.01$   & $0.12 \pm 0.01$   & $0.10 \pm 0.01$   & $ 0.05 \pm 0.01$ \\
$[\ion{O}{i}]\,63\,\mum$   & $0.24 \pm 0.01$  & $0.22 \pm 0.01$   & $0.28 \pm 0.01$   & $0.35 \pm 0.01$   & $ 0.35 \pm 0.01$ \\
$[\ion{O}{iii}]\,88\,\mum$ & $0.21 \pm 0.01$  & $0.23 \pm 0.01$   & $0.22 \pm 0.01$   & $0.18 \pm 0.01$   & $ 0.24 \pm 0.01$ \\
$[\ion{N}{ii}]\,122\,\mum$ & $0.30 \pm 0.01$  & $0.31 \pm 0.01$   & $0.28 \pm 0.01$   & $0.28 \pm 0.02$   & $ 0.28 \pm 0.01$ \\
$[\ion{O}{i}]\,146\,\mum$  & $0.01 \pm 0.01$  & $0.04 \pm 0.02$   & $0.03 \pm 0.01$   & $0.02 \pm 0.00$   & $ 0.05 \pm 0.01$  \\
$[\ion{C}{ii}]\,158\,\mum$ & $1.74 \pm 0.01$  & $1.64 \pm 0.02$   & $1.67 \pm 0.01$   & $1.72 \pm 0.02$   & $ 1.78 \pm 0.02$ \\
$[\ion{N}{ii}]\,205\,\mum$ & $0.15 \pm 0.01$  & $0.12 \pm 0.01$   & $0.15 \pm 0.01$   & $0.20 \pm 0.01$   & $ 0.20 \pm 0.01$  \\
\hline
&  spaxel (3,0) & spaxel (3,1) & spaxel (3,2) & spaxel (3,3) & spaxel (3,4)\\
\hline
$[\ion{N}{iii}]\,57\,\mum$ & $0.06 \pm 0.01$  & $0.01 \pm 0.01$   & $0.08 \pm 0.01$   & $0.03 \pm 0.01$   & $ 0.11 \pm 0.01$ \\
$[\ion{O}{i}]\,63\,\mum$   & $0.24 \pm 0.01$ & $0.27 \pm 0.02$  & $0.25 \pm 0.01$   & $0.28 \pm 0.01$   & $ 0.35 \pm 0.01$ \\
$[\ion{O}{iii}]\,88\,\mum$ & $0.21 \pm 0.01$  & $0.18 \pm 0.01$   & $0.20 \pm 0.01$   & $0.19 \pm 0.01$   & $ 0.26 \pm 0.01$ \\
$[\ion{N}{ii}]\,122\,\mum$ & $0.26 \pm 0.01$ & $0.32 \pm 0.01$  & $0.24 \pm 0.01$   & $0.28 \pm 0.01$   & $ 0.31 \pm 0.01$ \\
$[\ion{O}{i}]\,146\,\mum$  & $0.02 \pm 0.01$ & $0.01 \pm 0.01$  & $0.02 \pm 0.02$   & $0.03 \pm 0.01$   & $ 0.03 \pm 0.01$  \\
$[\ion{C}{ii}]\,158\,\mum$ & $1.68 \pm 0.02$ & $1.63 \pm 0.01$  & $1.63 \pm 0.02$   & $1.71 \pm 0.01$   & $ 1.76 \pm 0.01$ \\
$[\ion{N}{ii}]\,205\,\mum$ & $0.17 \pm 0.01$ & $0.16 \pm 0.01$  & $0.14 \pm 0.01$   & $0.12 \pm 0.01$   & $ 0.08 \pm 0.01$  \\
\hline
&  spaxel (4,0) & spaxel (4,1) & spaxel (4,2) & spaxel (4,3) & spaxel (4,4)\\
\hline
$[\ion{N}{iii}]\,57\,\mum$ & $0.12 \pm 0.01$  & $0.10 \pm 0.00$   & $0.05 \pm 0.00 $   & $0.08 \pm 0.00 $   & $ 0.08 \pm 0.00 $ \\
$[\ion{O}{i}]\,63\,\mum$   & $0.25 \pm 0.01$ & $0.26 \pm 0.01$  & $0.29 \pm 0.01$   & $0.30 \pm 0.01$   & $ 0.29 \pm 0.00$ \\
$[\ion{O}{iii}]\,88\,\mum$ & $0.20 \pm 0.01$  & $0.23 \pm 0.01$   & $0.22 \pm 0.01 $   & $0.17 \pm 0.01 $   & $ 0.22 \pm 0.00 $ \\
$[\ion{N}{ii}]\,122\,\mum$ & $0.25 \pm 0.01$ & $0.29 \pm 0.01$  & $0.43 \pm 0.01$   & $0.22 \pm 0.00$   & $ 0.26 \pm 0.01$ \\
$[\ion{O}{i}]\,146\,\mum$  & $0.01 \pm 0.01$ & $0.02 \pm 0.02$  & $0.04 \pm 0.01$   & $0.04 \pm 0.00$   & $ 0.02 \pm 0.01$  \\
$[\ion{C}{ii}]\,158\,\mum$ & $1.61 \pm 0.02$ & $1.59 \pm 0.02$  & $1.58 \pm 0.02$   & $1.59 \pm 0.01$   & $ 1.58 \pm 0.01$ \\
$[\ion{N}{ii}]\,205\,\mum$ & $0.10 \pm 0.01$ & $0.10 \pm 0.01$  & $0.14 \pm 0.01$   & $0.17 \pm 0.00$   & $ 0.17 \pm 0.01$  \\
\hline                                   
\end{tabular}
\end{table*}

\end{appendix}

\end{document}